\documentclass[twocolumn]{aastex631}

\shorttitle{ICL measurement with machine learning}
\shortauthors{Canepa et al.}
\begin{document}

\title{Measuring the intracluster light fraction with machine learning}

\author{Louisa Canepa}
\affiliation{School of Physics, University of New South Wales, NSW 2052, Australia}
\affiliation{ARC Centre of Excellence for All Sky Astrophysics in 3 Dimensions (ASTRO 3D)}

\author{Sarah Brough}
\affiliation{School of Physics, University of New South Wales, NSW 2052, Australia}
\affiliation{ARC Centre of Excellence for All Sky Astrophysics in 3 Dimensions (ASTRO 3D)}

\author{Francois Lanusse}
\affiliation{The Flatiron Institute, 162 5th Ave, New York, NY 10010, USA}
\affiliation{Universit\'e Paris-Saclay, Universit\'e Paris Cit\'e, Paris 91190, France}

\author{Mireia Montes}
\affiliation{Institute of Space Sciences (ICE, CSIC), Campus UAB, Carrer de Can Magrans, s/n, 08193 08193, Cerdanyola del Valles, Spain}
\affiliation{Instituto de Astrof\'isica de Canarias, Calle Vía Láctea s/n, 38204, San Crist\'obal de La Laguna, Tenerife, Spain}
\affiliation{Departamento de Astrof\'isica, Universidad de La Laguna, 38206, La Laguna, Tenerife, Spain}

\author{Nina Hatch}
\affiliation{School of Physics and Astronomy, University of Nottingham, University Park, Nottingham NG7 2RD, UK}

%% Note that the \and command from previous versions of AASTeX is now
%% depreciated in this version as it is no longer necessary. AASTeX 
%% automatically takes care of all commas and "and"s between authors names.

%% AASTeX 6.31 has the new \collaboration and \nocollaboration commands to
%% provide the collaboration status of a group of authors. These commands 
%% can be used either before or after the list of corresponding authors. The
%% argument for \collaboration is the collaboration identifier. Authors are
%% encouraged to surround collaboration identifiers with ()s. The 
%% \nocollaboration command takes no argument and exists to indicate that
%% the nearby authors are not part of surrounding collaborations.

%% Mark off the abstract in the ``abstract'' environment. 
\begin{abstract}

The intracluster light (ICL) is an important tracer of a galaxy cluster's history and past interactions. However, only small samples have been studied to date due to its very low surface brightness and the heavy manual involvement required for the majority of measurement algorithms. Upcoming large imaging surveys such as the Vera C. Rubin Observatory's Legacy Survey of Space and Time are expected to vastly expand available samples of deep cluster images. However, to process this increased amount of data, we need faster, fully automated methods to streamline the measurement process. This paper presents a machine learning model designed to automatically measure the ICL fraction in large samples of images, with no manual preprocessing required. We train the fully supervised model on a training dataset of 50,000 images with injected artificial ICL profiles. We then transfer its learning onto real data by fine-tuning with a sample of 101 real clusters with their ICL fraction measured manually using the surface brightness threshold method. With this process, the model is able to effectively learn the task and then adapt its learning to real cluster images. Our model can be directly applied to Hyper Suprime-Cam images, processing up to 500 images in a matter of seconds on a single GPU, or fine-tuned for other imaging surveys such as LSST, with the fine-tuning process taking just 3 minutes. The model could also be retrained to match other ICL measurement methods. Our model and the code for training it is made available on GitHub.

\end{abstract}

%% Keywords should appear after the \end{abstract} command. 
%% The AAS Journals now uses Unified Astronomy Thesaurus concepts:
%% https://astrothesaurus.org
%% You will be asked to selected these concepts during the submission process
%% but this old "keyword" functionality is maintained in case authors want
%% to include these concepts in their preprints.
% \keywords{Classical Novae (251) --- Ultraviolet astronomy(1736) --- History of astronomy(1868) --- Interdisciplinary astronomy(804)}

%% From the front matter, we move on to the body of the paper.
%% Sections are demarcated by \section and \subsection, respectively.
%% Observe the use of the LaTeX \label
%% command after the \subsection to give a symbolic KEY to the
%% subsection for cross-referencing in a \ref command.
%% You can use LaTeX's \ref and \label commands to keep track of
%% cross-references to sections, equations, tables, and figures.
%% That way, if you change the order of any elements, LaTeX will
%% automatically renumber them.
%%
%% We recommend that authors also use the natbib \citep
%% and \citet commands to identify citations.  The citations are
%% tied to the reference list via symbolic KEYs. The KEY corresponds
%% to the KEY in the \bibitem in the reference list below. 

\section{Introduction} \label{sec:intro}

Galaxy groups and clusters contain a diffuse, low surface brightness component that extends throughout the cluster (e.g. \citealp{mihos_diffuse_2005}). This component is known as the intracluster light (ICL; see \citealp{mihos_deep_2019}; \citealp{contini_origin_2021}; \citealp{montes_faint_2022} for reviews), and is thought to be mainly made up of stars that have been ripped out of their original galaxies during the accretion history of the cluster, through mergers and interactions with other galaxies (e.g. \citealp{rudick_formation_2006}; \citealp{contini_formation_2014}). In-situ star formation might also be a contributor, albeit less significant, to the ICL (e.g. \citealp{barfety_assessment_2022}; \citealp{montenegro-taborda_growth_2023}). The ICL forms an important record of the formation history of the cluster and the evolution of the galaxies within it, particularly the brightest cluster galaxy (BCG), around which the ICL is concentrated (e.g. \citealp{rudick_quantity_2011}; \citealp{canas_stellar_2020}). 

One metric that is widely used to quantify the amount of ICL in a cluster is the ICL fraction, that is, the fraction of the total stellar light in the cluster that belongs to the ICL. The relationship of this quantity with host cluster properties such as redshift, mass, or dynamical state can provide key insights into the efficiency of interactions in different environments. 

Simulations suggest an increasing amount of ICL over time (with decreasing redshift), as more stars are added to the ICL as the cluster evolves (e.g. \citealp{murante_importance_2007}). However, this relationship could be non-monotonic, with accretion of other groups and clusters with lower ICL fractions causing the ICL fraction of the combined system to decrease, while disruptions of the individual galaxies already within the cluster cause the fraction to increase \citep{canas_stellar_2020}. The overall relationship is then the result of the interplay between the timescales of these processes. Simulations disagree about the expected strength of this relationship however, with \cite{montes_faint_2022} showing that, for example, \cite{rudick_quantity_2011} found a much more gradual increase across redshift than \cite{tang_investigation_2018}. 

Some simulations suggest a weakly increasing correlation between ICL fraction and host cluster halo mass in the group-cluster mass range (e.g. \citealp{purcell_shredded_2007}; \citealp{canas_stellar_2020}; \citealp{proctor_identifying_2024}), albeit with significant scatter, while others do not find any trend (e.g. \citealp{rudick_quantity_2011}; \citealp{contini_formation_2014}; \citealp{contreras-santos_characterising_2024}). 

Simulations also suggest a relationship between the amount of ICL in a cluster and the cluster's dynamical state, where relaxed and hence evolved clusters have higher ICL fractions than their counterparts, although noting that simulations define the ICL fraction in mass rather than stellar light(e.g. \citealp{contini_intracluster_2023}; \citealp{brough_preparing_2024}; \citealp{contreras-santos_characterising_2024}). 

Analyzing these trends with ICL fraction and cluster properties using observational data presents difficulties. The observational definition of the ICL is ambiguous, and it is unclear how to best separate the light in the ICL from the rest of the light in the cluster, in particular the BCG. There are a number of different methods currently used. These range from using a surface brightness threshold to define and separate the ICL (e.g. \citealp{feldmeier_deep_2004}; \citealp{burke_coevolution_2015}; \citealp{montes_intracluster_2018}; \citealp{furnell_growth_2021}; \citealp{martinez-lombilla_galaxy_2023}), to modeling and subtracting cluster galaxies through either analytical models (e.g. \citealp{gonzalez_intracluster_2005}; \citealp{spavone_vegas_2017}; \citealp{morishita_characterizing_2017}; \citealp{ragusa_vegas_2021}) or orthonormal mathematical bases (e.g. \citealp{jimenez-teja_disentangling_2016}; \citealp{jimenez-teja_dissecting_2023}), to using wavelet-based algorithms (e.g. \citealp{da_rocha_intragroup_2005}; \citealp{guennou_intracluster_2012}; \citealp{adami_diffuse_2013}; \citealp{ellien_dawis_2021}). These each come with their own assumptions and biases, which makes it particularly challenging to compare between different existing samples and study trends on a large scale. 

The review by \cite{montes_faint_2022} compared measurements taken with the surface brightness threshold method, composite model method, and wavelet method. They found that the observed relationship between ICL fraction and redshift was strongly dependent on the measurement method chosen, with the surface brightness threshold method showing an increase in ICL fraction with decreasing redshift, the composite model method showing slight evolution in redshift, and the final method showing no correlation. There was also significant scatter between different samples due to the different depths, bands, and surface brightness thresholds applied in the images. These issues make it difficult to compare with expectations from simulations and to compare different observations with one another. However, with the heterogeneous samples available, observations also seem to indicate that there is no trend of ICL fraction with host halo mass (e.g. \citealp{krick_diffuse_2007}; \citealp{montes_faint_2022}; \citealp{ragusa_does_2023}). Finally, there are also contradictory results from observations regarding the correlation with cluster relaxation -- some studies (e.g. \citealp{da_rocha_intragroup_2008}; \citealp{iodice_vegas_2020}; \citealp{ragusa_does_2023}) found that the ICL fraction appears to be higher in more dynamically evolved systems, whereas \cite{jimenez-teja_unveiling_2018} find the opposite trend. These results show the difficulty posed by drawing conclusions from heterogeneous samples. 

\cite{brough_preparing_2024} compared eight different observational measurement methods applied to a homogeneous sample of simulations. They found agreement between the mean ICL fraction measured by the different methods, but found significant scatter present on an individual measurement basis. The largest difference in ICL fraction for a single cluster was 23\% (ranging from 0.11 to 0.34). These results indicate that using a homogeneous sample does alleviate some of the difficulties in comparison, and that the measurement methods do show broad agreement, however the scatter involved still makes direct comparison between samples analyzed with different methods challenging. 

These issues are compounded by the dependence of the ICL fraction on the photometric band and survey limiting depth used for the observation (e.g. \citealp{burke_coevolution_2015}; \citealp{montes_intracluster_2018}). Constructing a large, homogeneous sample of ICL measurements using a single method would be the most effective way to remove these systematic issues and draw more robust conclusions about the ICL fraction's dependence on host cluster properties. 

To generate such a sample, we need many homogeneous cluster images with the depth required to study this low surface brightness component. With the next generation of optical imaging surveys expected to reach unprecedented depths, we will soon have enough data to solve this issue. The Vera C. Rubin Observatory's Legacy Survey of Space and Time (LSST; \citealp{ivezic_lsst_2019}) will image the entire Southern sky to a depth sufficient to study the ICL (\citealp{montes_intracluster_2019}; \citealp{brough_vera_2020}). Thus, to take full advantage of these data and avoid the problems presented by different measurement methods, we need to be able to efficiently measure the ICL fraction in hundreds of thousands of group and cluster images, as LSST is likely to find $\sim$100,000 clusters and $\sim$1 million groups up to $z\sim1.2$ over the southern hemisphere \citep{brough_vera_2020}. However, the measurement methods currently available are not readily scalable to sample sizes of this magnitude, as many of them rely on some manual involvement and tuning on a per-cluster basis (e.g. \citealp{brough_preparing_2024}). Alongside investigation into making these methods applicable to significantly larger samples, we need to explore other possible options. 

Machine learning has proven to be particularly useful for these types of data-intensive problems, as it requires little to no human involvement, making it an attractive solution for effectively processing large numbers of images. Machine learning models have been developed for a wide array of problems in astronomy, such as image denoising (e.g. \citealp{vojtekova_learning_2021}) and outlier identification (e.g. \citealp{han_identifying_2022}). In the ICL field, \citet{marini_machine_2022} used a machine learning model to classify star particles in simulated galaxy clusters as belonging to either the ICL or the BCG based on the given host cluster mass, normalized particle cluster-centric distance, and rest-frame velocity. Machine learning models have seen particular use in image classification problems, for example identification of galaxy mergers \citep{pearson_identifying_2019}, morphological classification of galaxies \citep{hayat_self-supervised_2021}, and classification of tidal features \citep{desmons_detecting_2024}. They have also seen use in image regression problems, for example in predicting photometric redshifts from input images \citep{hayat_self-supervised_2021}.

A difficulty presented in astronomy applications for machine learning, in particular ICL measurement, is the lack of labeled training samples. Supervised machine learning models often require tens of thousands of training samples in order to learn a task sufficiently well, and such datasets do not exist even for current ICL measurement methods. Some works (e.g. \citealp{hayat_self-supervised_2021}; \citealp{stein_mining_2022}; \citealp{desmons_detecting_2024}) solve this problem by using self-supervised representation learning, which uses a combination of a large unlabeled dataset and a small labeled dataset to learn a task. However, this technique is not currently suitable for ICL measurement, owing to the limited numbers of deep images of clusters available to construct a large enough unlabeled dataset. Instead, here we explore a technique called transfer learning, which involves adapting learning from a different task or dataset to the target domain \citep{pan_survey_2010}. It is a well known issue that machine learning models struggle with different datasets. For example, \citet{pearson_identifying_2019} found that the accuracy of a model trained on real images fell from 91.5\% to 53.0\% when applied to simulated mock images. However, if the two datasets are similar enough, this issue can be mitigated with fine-tuning (e.g. \citealp{oquab_learning_2014}; \citealp{walmsley_practical_2022}). This involves taking a model pretrained on a very large dataset related to the target task, and training it further on a small sample of images from the target dataset. Generally, either only the final few layers are trained or a low learning rate is used, to avoid destroying all the learned weights from the previous task. 

In this work, we develop a machine learning model to efficiently measure the ICL fraction in input cluster images. Section \ref{sec:method} describes our data sources, outlines the methods used to construct our training and fine-tuning datasets, and describes the architecture of the machine learning model. Section \ref{sec:results} presents the results of our model, and Section \ref{sec:discussion} discusses the reliability of the results and potential areas of improvement for the model. Throughout this work we assume a standard cosmological model with parameters $H_0=70$ km s$^{-1}$ Mpc$^{-1}$, $\Omega_m=0.3$, and $\Omega_\Lambda=0.7$.

\section{Method}\label{sec:method}
\subsection{Data sources}
We use images from the Hyper Suprime-Cam Subaru Strategic Program (HSC-SSP; \citealp{miyazaki_hyper_2018}) Public Data Release 2 (PDR2; \citealp{aihara_second_2019}). Although Public Data Release 3 (PDR3; \citealp{aihara_third_2022}) is now available, we choose to use PDR2 as it fulfils the requirements of our work and has been extensively tested for low surface-brightness studies (e.g. \citealp{huang_individual_2018}; \citealp{huang_weak_2020}; \citealp{li_reaching_2022}) whereas PDR3 introduces several differences in sky conditions, observing strategy, and data treatment pipeline that would need to be accounted for. HSC-SSP is carried out with the Hyper Suprime-Cam instrument on the 8.2m Subaru telescope, and is a three layered multi-band (\textit{grizy}) imaging survey. The three layers have varying limiting depths: Wide ($m_r\sim26.4$ mag), Deep ($m_r\sim27.4 $mag), and Ultradeep ($m_r\sim28.0$ mag). LSST is expected to image the entire southern sky to a surface brightness depth\footnote{https://smtn-016.lsst.io/} of $\mu_r \sim 30.3$ mag/arcsec$^2$ (3$\sigma$, $10\arcsec\times10\arcsec$), which is comparable to the Ultradeep layer of HSC, which has a depth of $\mu_r \sim 29.8$ mag/arcsec$^2$ \citep{martinez-lombilla_galaxy_2023}. The Deep and Ultradeep layers are imaged to a depth sufficient to study low surface brightness features such as ICL, and also has the advantage that it uses the same pipeline as LSST to reduce its data \citep{bosch_hyper_2018}, meaning that a model trained on HSC-SSP data should be easily adapted to LSST data as well. This is important as the need for an efficient measuring algorithm such as the one we present here is driven by the amount of data that LSST will provide. 

In order to make manual measurements of the ICL, we require a catalogue of clusters. We use CAMIRA \citep{oguri_optically-selected_2018}, an optically-selected catalogue of clusters run on all HSC-SSP images. The algorithm, fully described in \citet{oguri_cluster_2014}, leverages the expectation that all massive clusters are expected to show a ``red sequence" of galaxies (e.g. \citealp{gladders_new_2000}) to identify likely red sequence galaxies using a stellar population synthesis model and group them into clusters. 

The final product of the CAMIRA algorithm is a catalogue of clusters, where the central coordinates of each cluster are the coordinates of the BCG identified by the algorithm. We only use the Deep/Ultradeep clusters for our manual ICL measurements, as the depths of these layers are best suited for low surface brightness studies. The Wide clusters are used to define a BCG stellar mass distribution, which is used in the construction of our training sample as described below. In the Deep/Ultradeep layers 248 clusters are identified by the CAMIRA algorithm, and 7939 clusters are identified in the Wide layer. Photometric cluster redshifts are calculated by the CAMIRA algorithm based on the most likely redshift for a particular likely red sequence of galaxies, as described in detail in \citet{oguri_cluster_2014}. We restrict our sample to clusters that have a photometric redshift of less than 0.5, as low surface brightness features become extremely challenging to detect past this limit due to cosmological dimming. 130 clusters in the Deep/Ultradeep layers and 2709 clusters in the Wide layer fit this criterion. These clusters were identified in the HSC-SSP PDR3 survey, which covers more area than the PDR2 that we use for our image data (1298 square degrees compared to 1022 square degrees in the Wide layer, and 36 square degrees compared to 35 square degrees in the Deep/Ultradeep layers). Using PDR2 images reduces our sample size to 129 clusters in the Deep layer and 2379 clusters in the Wide layer. We also remove clusters from these layers that have bad photometry or have significant contamination from bright stars (identified by manual inspection of these images), leaving us with 101 clusters in the Deep/Ultradeep layers. 

\begin{figure}
    \centering
    \includegraphics[width=0.95\linewidth]{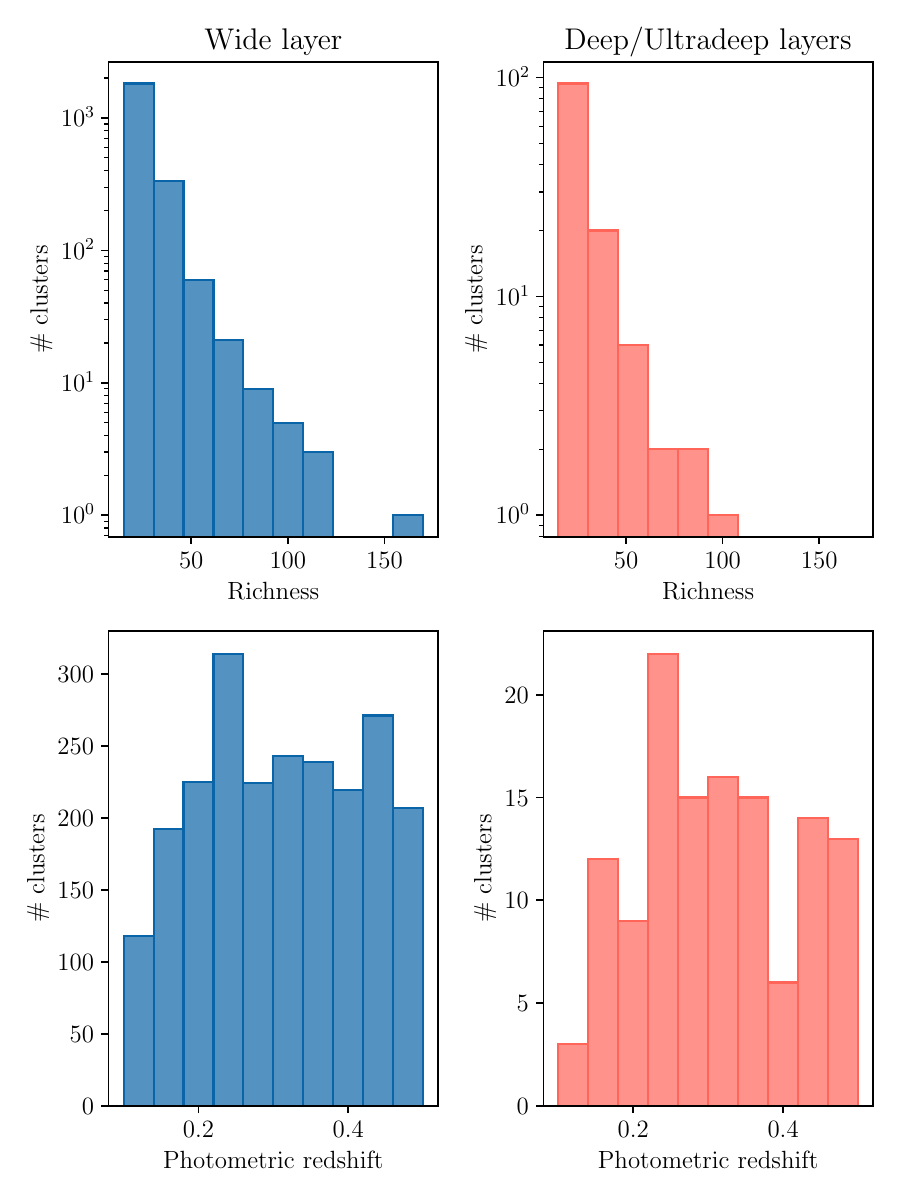}
    \caption{Distribution of cluster richness (top) and photometric cluster redshift (bottom) of the final Wide sample (left) and Deep/Ultradeep sample (right). Values are as provided by the CAMIRA algorithm.}
    \label{fig:bcg_distributions}
\end{figure}

The redshifts of these clusters have a range of $0.1<z<0.5$, where $0.5$ is our chosen upper limit as described previously and $0.1$ is the lower limit of the CAMIRA catalogue due to the large angular sizes and brightness of member galaxies in HSC-SSP below this redshift complicating the cluster finding \citep{oguri_optically-selected_2018}. The parameter richness in CAMIRA is defined as the number of red member galaxies with stellar masses $M_*\gtrsim10^{10.2}M_\odot$, within an aperture of radius $R\sim1h^{-1}$Mpc, where the stellar mass of all likely red sequence galaxies is predicted by the stellar population synthesis model \citep{oguri_cluster_2014}. In our final sample, the cluster richness ranges from 15 to 93 in the Deep/Ultradeep layers and 15 to 170 in the Wide layer. This lower limit roughly corresponds to $M_{200\textrm{m}} \gtrsim 10^{14}h^{-1}M_\odot$ \citep{oguri_optically-selected_2018}. The richness and redshift distribution of the samples over all three layers is shown in Figure ~\ref{fig:bcg_distributions}.

The CAMIRA algorithm also produces a catalogue of 2,191,333 galaxies from the Wide layer that fit the red sequence stellar population synthesis model well, called Luminous Red Galaxies (LRGs), which we also make use of in our work. As with the clusters, we exclude galaxies with redshifts higher than 0.5, leaving a sample of 479,435 LRGs. 

Finally, in this work we make use of the photometric redshifts provided by HSC-SSP \citep{nishizawa_photometric_2020}, calculated using the Mizuki template fitting code \citep{tanaka_photometric_2018}, which gives photometric redshifts for all objects in the Deep/Ultradeep layers of HSC-SSP.

\subsection{Training data}
Machine learning models require significant amounts of data (ideally tens of thousands of samples) to effectively learn a task. Given that we only have a sample of 2480 cluster images over the entire HSC-SSP survey, and only 101 of these in the Deep/Ultradeep layer, it is currently not possible to assemble such a sample for training. We use transfer learning to overcome this issue by generating a large sample of artificial images to use as our main training dataset, and then transfer the model's learning into the real domain using the smaller sample of real images. 

Transfer learning is most effective when the training domain is as similar as possible to the target domain. For this reason, rather than using simulated data, which would give us realistic ICL morphologies based on real physical processes and the interaction history of the cluster, we choose to prioritize the realistic appearance of the images by using existing HSC-SSP images as a base and injecting simple artificial ICL profiles. 

We use the LRGs identified by the CAMIRA algorithm in the Wide layer as our base images of galaxies. We use these instead of injecting many different artificial ICL profiles into repeat versions of the real BCG images to have more diversity in the training sample for more robust learning. Using the limited number of real BCGs in the Deep layers as the base images could lead to the model overfitting to those particular clusters, which would ultimately hurt its ability to generalize to unseen data. BCGs are typically a special case of the LRG population (e.g. \citealp{loh_bright_2006}; \citealp{brough_spatially_2007}; \citealp{dalal_brightest_2021}), meaning that the generated images, centered on LRGs, will look more similar to the real images, centered on BCGs, compared to a sample of randomly drawn galaxies.

\begin{figure}
    \centering
    \includegraphics[width=0.95\linewidth]{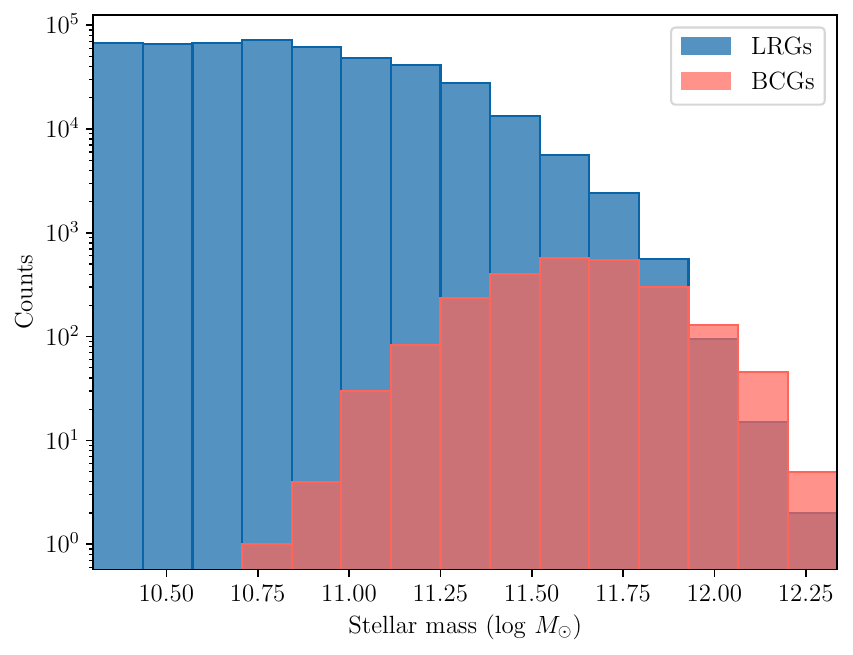}
    \caption{Stellar mass distribution of the Wide layer LRG sample with redshift $<0.5$, and the combined BCG sample in Wide, Deep, and Ultradeep layers.}
    \label{fig:mass_distributions}
\end{figure}

The stellar mass distribution of the LRG sample is significantly different to the combined sample of BCGs from all layers, as shown in Figure~\ref{fig:mass_distributions}. The reason that the number of BCGs in high mass bins exceeds the number of LRGs is that the BCG sample is drawn from all three layers, whereas the LRG sample is only from the Wide layer. This difference in the two distributions means that a random sample drawn from images of LRGs will contain many more low mass galaxies than the BCG sample. In order to keep our datasets as similar as possible, we re-sample the LRGs to match the mass distribution of our BCG sample. To do this, we estimate the probability density function (PDF) of the BCG mass distribution and the LRG mass distribution using Gaussian kernel density estimation. We then weight the probability of sampling each galaxy by the ratio of the BCG mass PDF to the LRG mass PDF at that galaxy's mass. In this way, the likelihood of drawing an LRG of the same mass as a BCG increases. We sample 50,000 LRGs using this method, as this should be a sufficient number for the supervised training. After this weighted sampling of the LRGs, the stellar mass distribution of our drawn sample of 50,000 LRGs matches the mass distribution of the BCGs.

We use $600\times600$ kpc cutouts in $r$-band from the Wide layer of the HSC-SSP survey, converting this into an angular size for downloading each cluster. Given the HSC pixel scale of 0.168 arcsec, this covers a range of 1936 pixels at $z=0.1$ to 585 pixels at $z=0.5$. Current observations generally detect ICL out to radii of between 100 kpc and 300 kpc depending on the survey, before it becomes too faint compared to the background level to be detected (e.g. \citealp[]{montes_intracluster_2014}; \citealp{demaio_origin_2015}; \citealp{jimenez-teja_unveiling_2018}; \citealp{montes_buildup_2021}). This physical size should therefore be large enough to capture the ICL in the majority of cases. We then mask out bright stars in the image using the bright star masks provided by the HSC-SSP survey.

\begin{figure*}
    \centering
    \includegraphics[width=0.95\linewidth]{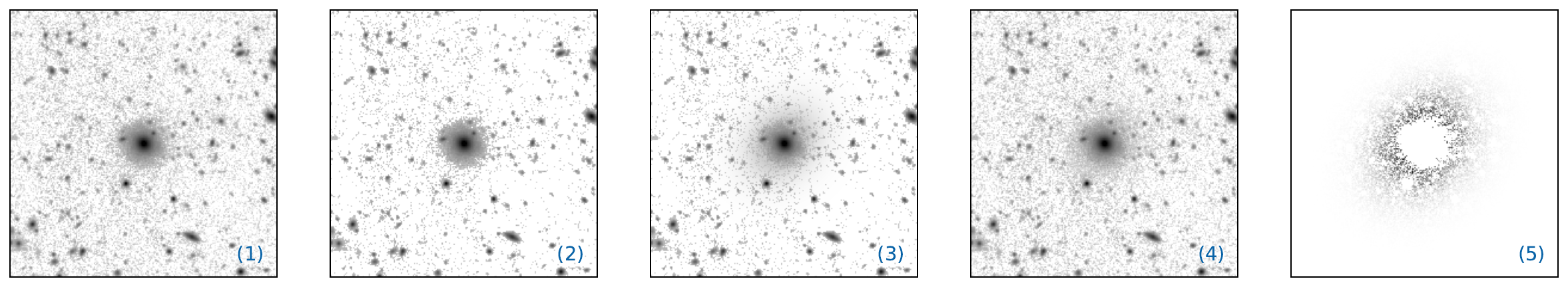}
    \caption{From left to right the images show the steps of the training data generation procedure. (1) The original cutout of an LRG from HSC-PDR2 wide data, resized to 224x224 pixels. (2) The image with light fainter than 26 mag/arcsec$^2$ removed. (3) Randomly generated exponential profile added to the thresholded image. (4) Final generated image with extra noise and background added. (5) The difference between images (1) and (4) to better highlight the added low surface brightness light. All images have been arcsinh scaled.}
    \label{fig:data_gen}
\end{figure*}

The method of data generation is illustrated in Figure~\ref{fig:data_gen}. Firstly, we remove all light in the image fainter than a surface brightness of $\mu_r=26$ mag/arcsec$^2$ (e.g. \citealp{rudick_quantity_2011}; \citealp{montes_intracluster_2018}; \citealp{martinez-lombilla_galaxy_2023}), corrected for cosmological dimming and $k$-corrected following Equation~\ref{eq:sb_threshold}, where the $k$-correction, $k(z)$, is estimated as a function of redshift using the ``k-corrections calculator"\footnote{http://kcor.sai.msu.ru/} \citep{chilingarian_analytical_2010} and assuming an ICL color $g-r = 0.7$ \citep{martinez-lombilla_galaxy_2023}. 
\begin{equation}
    \mu_{obs}(z) = 26 + 2.5 \log_{10}(1+z)^4 + k(z)
    \label{eq:sb_threshold}
\end{equation}

This removes any existing low surface brightness features around the central galaxy, including any potential ICL (see panel 2 in Figure~\ref{fig:data_gen}).

We then randomly generate an artificial ICL profile, using an exponential profile with randomized ellipticity and position angle (c.f. \citealp{watkins_strategies_2024}). We apply a surface brightness cut to find the 26 mag/arcsec$^2$ isophote of the LRG, and set the effective radius of the profile to a randomly selected pixel size between the semi-major and semi-minor axes of the isophote. Several studies have found that the ICL can be modeled using an exponential profile (e.g. \citealp{montes_intracluster_2018}; \citealp{kluge_photometric_2021}; \citealp{ragusa_vegas_2021}). We visually determined that our parameters automatically produce a reasonable ICL profile for a number of different LRG base images. We also tested the impact of modeling ICL with an exponential profile by training a model on images injected with S\'ersic profiles with random indices between 1 and 8. We find that this does have a minor impact on the model’s performance, with a Mean Absolute Error (MAE) in its predictions of 0.0186 compared to our final model’s MAE of 0.0159. This means that for an ICL fraction of 0.2, the error on the fraction increases from 7.9\% to 9.3\%. This indicates that even using wildly unrealistic profiles during initial model training does not have a significant impact on the model’s ability to recognize the ICL profiles in real images. However, we note that the use of this profile could create the possibility that our model may not be as adept at identifying ICL with more asymmetric morphologies (e.g. \citealp{rudick_formation_2006}; \citealp{presotto_intracluster_2014}; \citealp{montes_new_2022}). We then inject this profile into the thresholded image (see panel 3 in Figure~\ref{fig:data_gen}). 

In the first step, we removed all low surface brightness light in the image. This is intended to remove any low surface brightness light initially present around the central galaxy, which is replaced by our injected ICL profile. However, this will also remove any other low surface brightness light in the image associated with other sources. Removing this light gives these other sources an unnatural appearance that could hinder the model's ability to process real images. To test the impact of this step, we trained a model on images that instead have Gaussian noise injected as the background. We find that this causes the model to fail to make predictions on real data. This shows that maintaining the realistic appearance and noise properties of the images is important to enabling the model to transfer its learning to the target dataset. We therefore want to preserve this light in our generated training images. To do this, we create a version of the ICL profile which is normalized between 0 and 1 (where 1 corresponds to the maximum flux value, equivalent to 26 mag/arcsec$^2$). This is then inverted, and used to scale the original subtracted background before adding it to the image. This means that where the ICL profile is close to 0, away from the central galaxy, the image has its original background, but as the profile gets brighter, the original background is suppressed, leaving our injected ICL profile around the central galaxy. Finally, so that the ICL does not appear unnaturally noiseless as in panel 3 of Figure~\ref{fig:data_gen}, we calculate the $3\sigma$ clipped statistics of the original background, and add Gaussian noise with the same standard deviation to the image, scaled by the normalized ICL profile so that the noise across the new image remains consistent with the original. This new image is the final generated image (see panel 4 in Figure~\ref{fig:data_gen}). 

Running this process on all 50,000 images of LRGs gives us our final set of images for the initial stage of the model training. We refer to this dataset as the ``artificial dataset", in contrast to the ``fine-tuning dataset" which consists of the sample of 101 real Deep/Ultradeep cluster images that we introduce below. 

\subsection{Measurement method}
In order to train the model to predict the ICL fraction, we first need to measure the ICL fractions in both our artificial dataset and the fine-tuning dataset. There are multiple different possible methods to choose from to measure the ICL as discussed in \citet{montes_faint_2022} and \citet{brough_preparing_2024}. In this case we use the surface brightness cut method to make these measurements. This method assumes that all flux fainter than a chosen surface brightness threshold is contributed by the ICL component. \cite{brough_preparing_2024} found that this method gave results consistent with other measurement methods. We choose this method for its relative simplicity and the fact that it makes no assumptions about the morphology of either the BCG or the ICL. 

We measure the ICL using the surface brightness cut method for both our artificial and fine-tuning datasets, and describe our procedures in more detail below. The exact procedure differs slightly between the two. For the fine-tuning dataset, we manually tune the background subtraction and bright star masking, and also apply masks for non-cluster member galaxies on an individual cluster basis when conducting the measurement. The fine-tuning dataset is small enough that manual inspection and tuning of each cluster is feasible, and it is important to have accurate measurements for this dataset to achieve a good transfer of the model's learning into the final image domain. However, for the artificial dataset, we prioritize making the measurements completely automatically, as it is not feasible to manually tweak the measurement for 50,000 clusters. The relative simplicity of the method makes it easy to automate the process to some extent, but it is quite reliant on effective masking and background subtraction, which often requires human evaluation and involvement and makes it problematic to apply automatically. Also, since the artificial dataset does not contain actual clusters, we cannot apply cluster membership information here. This means that the automatic measurements taken of the artificial dataset are likely to be slightly inaccurate for some clusters due to these factors. However, given that the purpose of the artificial dataset is to give the model only a general understanding of the task, and that inaccuracies in its understanding should be corrected in the fine-tuning stage, this level of noise in the measurements should be acceptable for our artificial dataset. We explore the impact of this assumption in the next section.

\subsubsection{Measurement method for artificial dataset}
The model accepts images of size $224\times224$ pixels. This is a standard size for computer vision models, to lower the computational cost of training and inference, so we first resize the cutouts to this size. This is done using \textsc{scikit-image} \citep{van_der_walt_scikit-image_2014}, which interpolates the image to downsize the image. This resizing does mean that lower redshift clusters will be more heavily binned than higher redshift clusters due to their larger initial size, which could affect the measured ICL fraction. We return to this in Section \ref{sec:discussion}.

We then perform an automatic background estimate and subtraction. The HSC-SSP data has already been coadded and sky subtracted using a global sky subtraction algorithm \citep{aihara_second_2019}, however this often requires some additional correction, particularly to correct for gradients across the cutouts which would be problematic for ICL measurement. To estimate the background, we divide the cutout into a $14\times14$ grid, and use \textsc{Photutils} \citep{bradley_astropyphotutils_2023} to create a coarse 2D background map using the $3\sigma$ clipped statistics in each box, interpolating over bright stars. This is likely to include any ICL and the BCG in the center of the image as part of the background, as they extend over large areas in the image. To avoid this, we keep only the background values from boxes around the edge of the image, removing the central region of the coarse background map. We first calculate the $3\sigma$ clipped median of the counts in these boxes and subtract this constant sky value. We then fit a bivariate B-spline surface to these edge background values and use this as an estimate of the background. This avoids including any large features such as ICL in the center of the image as part of the background, and will smooth over any remaining sharp peaks in the background values due to extended galaxies on the edge of the image, while still allowing us to capture any large-scale 2D gradients across the image. We then subtract this estimate of the background from the image. As mentioned above, it is possible that this automatic procedure could lead to some over or under-subtraction in some images, leading to an increase in the uncertainties. However our procedure aims to minimize these issues as much as possible.

To test the sensitivity of the model’s learning to the choice of background subtraction procedure, we trained a model on measurements made with only a constant sky value subtracted, and another model with only the B-spline curve background estimate. After fine-tuning, the performance of these models was not significantly different from that of our final model, with an MAE of 0.0164 and 0.0158 compared to the final MAE of 0.0159. This indicates that the choice of background subtraction procedure in the artificial dataset does not have a strong effect on the model’s performance.

Next, we locate and mask very small background galaxies that might otherwise fall below the surface brightness threshold and artificially inflate the ICL fraction. We do this by unsharp masking the original image to increase the contrast before segmenting the image to create the ``hot mask", as done in \cite{montes_buildup_2021} and \cite{martinez-lombilla_galaxy_2023}. First, we mask out the parts of the image that are brighter than the $\mu_r=26$ mag/arcsec$^2$ threshold, as our goal here is to capture very small scale peaks or asymmetries fainter than this threshold that could be caused by background galaxies. The unsharp masking is then done by convolving the masked image with a Gaussian filter of $\sigma=2$ pixels using \textsc{Astropy} \citep{astropy_collaboration_astropy_2022}, subtracting it from the original, and segmenting the resulting image using \textsc{Photutils} methods. This will identify faint, small objects in the image that we then mask for the ICL calculation.

We then estimate the $3\sigma$ surface brightness limits of the image. We calculate the standard deviation of the sky by calculating the $3\sigma$ clipped standard deviation in the same region around the edges of the image from which we derived the background estimate, again to avoid including any extended light in the center of the image in our calculation. We then use this to find the surface brightness limit of the image as in Equation \ref{eq:sb_lim} following \cite{roman_galactic_2020}, where $\mu_\textrm{lim}(3\sigma_{10^{\prime\prime}\times10^{\prime\prime}})$ is the $3\sigma$ surface brightness limit for an angular scale of $10\times10$ arcsecond$^2$, $\sigma$ is our calculated standard deviation of the sky, pix is the pixel scale, and $ZP$ is the zero point of the image given by the HSC-SSP catalogue. 

\begin{equation}
\mu_\textrm{lim}(3\sigma_{10^{\prime\prime}\times10^{\prime\prime}}) = -2.5\times\log\Big(\frac{3\sigma}{\textrm{pix}\times 10}\Big)+ZP  \label{eq:sb_lim}
\end{equation}

Any pixels that fall below this calculated surface brightness limit are masked out to minimize background contamination in our calculation.

The ICL fraction is the flux from the ICL divided by the total flux from all galaxies and ICL in the cluster. However, the images from our artificial dataset are centered on LRGs rather than BCGs. So, for the artificial dataset measurements, we simplistically assume that all unmasked objects within the field of view belong to the cluster (which does not include bright stars or the small faint background galaxies as these have already been automatically masked). This is likely to lead to an overestimate of the cluster flux and a consequent underestimate of the ICL fraction, however, we leave the correction of this potential bias to the fine-tuning process. To investigate the effect of this assumption, we made a set of test measurements masking all galaxies other than the central LRG in the image, that is, assuming that the central LRG is the only cluster member, and a second set of test measurements where we classify galaxies with a photometric redshift within 3$\sigma$ of the central LRG as member galaxies (the same method that we use for our real BCGs, detailed below). The first set of test measurements results in worse performance after fine-tuning (an MAE of 0.0242 compared to 0.0159), because the model is trained to ignore satellite galaxies entirely during the initial training phase and this is too large of a difference for it to overcome during fine-tuning. On the other hand, the second test results in similar performance after fine-tuning (MAE of 0.0185 compared to our final MAE of 0.0159).

We choose a surface brightness threshold of $\mu_r = 26$ mag/arcsec$^2$, $k$-corrected and corrected for cosmological surface brightness dimming as for the data generation.

We apply the HSC-SSP provided bright star mask and the hot mask described above to the image. The ICL flux is calculated by summing the unmasked pixels below the chosen surface brightness threshold within a circular 300kpc radius aperture, covering the full radius of the image, which as previously described, should be large enough to capture the measurable extent of the ICL above the background in most cases. The total flux of the cluster is the sum of all unmasked pixels within the aperture. The ratio of these is then the final ICL fraction, which we use to label each of our artificial dataset images. For the artificial dataset, the measured ICL fractions range from 0.001 to 0.261, with a mean of $0.075\pm0.034$. This distribution arises naturally from our data generation process which prioritizes the realistic appearance of the generated ICL profiles. 

\subsubsection{Measurement method for fine-tuning dataset}
The measurement method for the fine-tuning dataset is largely similar to the artificial dataset, but at each step we visually inspect the image and make manual adjustments as necessary. We note that although these manual adjustments are necessary for the manual measurement of the fine-tuning dataset to ensure accurate labeling of the images, the images provided to the machine learning model during training and fine-tuning are not adjusted manually, i.e. they are not sky subtracted and only apply the provided HSC-SSP bright star masks. The model thus naturally learns to ignore the background of the images when making its predictions, meaning that when the model is later applied to samples of images, background subtraction and manual masking of bright objects is not necessary. The images we use for the manual measurements are not resized to $224\times224$ pixels before measurement.

Firstly, we visually inspect the image to find any obvious problems with the bright object masks provided by HSC-SSP. At times, these masks are not large enough to fully cover bright stars, or there are stars that have not been masked out, so we manually create and apply new masks to the images in these cases. 
 
We perform the background estimate as for the artificial dataset. After visual inspection of the result and manually checking the radial surface brightness profile of the image after subtraction, we then make manual adjustments to the background subtraction process as needed, such as adjusting the estimated gradient over the image or subtracting additional constant sky. 

To calculate the total cluster light for this sample, we use the photometric redshifts for other objects in the cutout to construct a mask that includes only the galaxies belonging to that cluster. We classify a galaxy as a cluster member if its photometric redshift is within $3\sigma$ of the cluster's, where the $1\sigma$ accuracy of the photometric redshifts is $0.05(1+z_{\textrm{phot}})$ \citep{tanaka_photometric_2018}. After identifying non-cluster members, we use the \textsc{Photutils} segmentation map to create an initial mask covering these galaxies. We enlarge these non-cluster member masks by convolving them with a Gaussian kernel of size 0.3 pixels after finding this size kernel worked well for covering the faint outskirts of these sources at our image resolution. We apply this mask to the image, so that the non-cluster member flux is not included in either the estimation of the ICL or total cluster light. 

We then construct the hot mask and mask below the surface brightness limits of the image as described previously, and finally calculate the ICL fraction as for the artificial dataset. The observational uncertainty on the ICL fraction is often estimated from the sky variance. However, the impact of background subtraction and masking on ICL fraction are significantly greater than the impact of sky variance in the images (mean = 0.001). We therefore use these steps to estimate the uncertainties on the fractions by measuring the images multiple times, spanning the range of parameters used in the manual measurements. We then take the standard deviation in the resulting ICL fraction as the uncertainty on the manually measured fraction.

\subsection{Model architecture}
We build a model that accepts a single band $224\times224$ pixel image as input, and produces a prediction for the ICL fraction for that image as output. A primary goal of this method is to remove the need for manual involvement. For this reason, the images given to the model are exactly as downloaded from HSC-SSP, with the only processing done being to downsize the images to the required resolution and applying the bright star masks as provided by HSC-SSP. No background subtraction or additional masking is performed. 

The images are first standardized by arcsinh stretching the images, dividing them by the mean absolute deviation of the images, and then normalizing them between 0 and 1. This suppresses the highest and lowest intensities of the image, and compresses the range to make it easier for the neural network to handle the inputs. 

The architecture of the network consists of an augmenter, an encoder, and output layers to generate the final answer. 

The augmenter introduces small random perturbations to input images that have no effect on the images' labels, and is key to showing the model what features of the image are unimportant to the task and preventing the model from overfitting to the training dataset. We randomly apply three image augmentations:
\begin{itemize}
    \item Randomly flipping the image horizontally and/or vertically, each with 50\% probability.
    \item Adding random Gaussian noise to the image with $\sigma$ equal to the mean absolute deviation of the unstandardized images. 
    \item Randomly rotating the image by 0 (no rotation), 90, 180, or 270 degrees, each with equal probability.
\end{itemize}

We base our encoder on a ResNet-50 architecture \citep{he_deep_2016} followed by a global average pooling layer to produce a 256-dimensional vector representation of the input image. 

The output layers then use this vector to produce the final ICL prediction. We represent this answer as a probability density distribution over the possible ICL fractions. This is done by first passing the 256-dimensional representation through a dense hidden layer with 2048 neurons, and then another dense layer which generates 48 outputs. These outputs are used as parameters to the output distribution, which is a mixture of beta distributions. We choose beta distributions as they are bounded between 0 and 1, which is a useful constraint for this problem. The mode of this distribution will then be the most probable value of the ICL fraction for the corresponding input image, and we retain information about the model's confidence in this prediction through the full probability distribution. 

\subsection{Model training and fine-tuning}
The model is built and trained using Tensorflow \citep{tensorflow_developers_tensorflow_2023}. It aims to maximize the log likelihood of the correct ICL fraction under the output probability distribution. This is implemented as a negative log likelihood loss function using the Tensorflow Probability library \citep{dillon_tensorflow_2017}. 

The model is compiled with the Adam optimizer \citep{kingma_adam_2014} and is first trained on the 50,000 images in the artificial dataset using 90\% (45,000) for training and 10\% (5,000) for validation. The learning rate starts at $10^{-4}$ and is decreased using a learning rate scheduler to $10^{-5}$ when the validation loss has not decreased for 10 epochs. The training is then stopped once the validation loss has not decreased for 25 epochs. Our final model was trained for 180 epochs with these conditions, taking 21 hours and 45 minutes on a single Tesla V100-SXM2 GPU.

We then need to transfer the model's learning onto real data using the fine-tuning data. A conventional 80-20 split for training and testing here will result in a testing sample too small to be able to accurately judge the final performance of the model, given that our full sample only has 101 cluster images. We therefore perform 5-fold cross-validation by repeating the fine-tuning process five times, using different partitions of the full dataset for testing each time. The model is trained from the same initial checkpoint each time. This firstly allows us to verify that the performance of our fine-tuned model is stable across different sets of training and testing data, and secondly, by collating the results of all five fine-tuned models, allows us to evaluate the effect of fine-tuning and performance of the final models over a statistically significant sample of all 101 real clusters. 

For the fine-tuning stage, we freeze all but the final two dense layers in the network, meaning that only the parameters in these final layers are allowed to change during fine-tuning, while the more fundamental weights of the encoder are kept fixed. This is because the model should only need small adjustments to transfer its learning, and allowing fewer parameters to vary will reduce the risk of the model overfitting to the small fine-tuning dataset. The model is then trained for 100 epochs at a learning rate of $10^{-6}$, which takes 3 minutes.

\section{Results}\label{sec:results}
In this section we first show the model's performance on the artificial dataset after the initial training stage, showing that it has successfully learned the ICL fraction prediction task for this first dataset. We then compare the model's performance on the real data before and after the fine-tuning step to explore the effect of fine-tuning.

\subsection{Artificial dataset}
\begin{figure}
    \centering
    \includegraphics[width=0.95\linewidth]{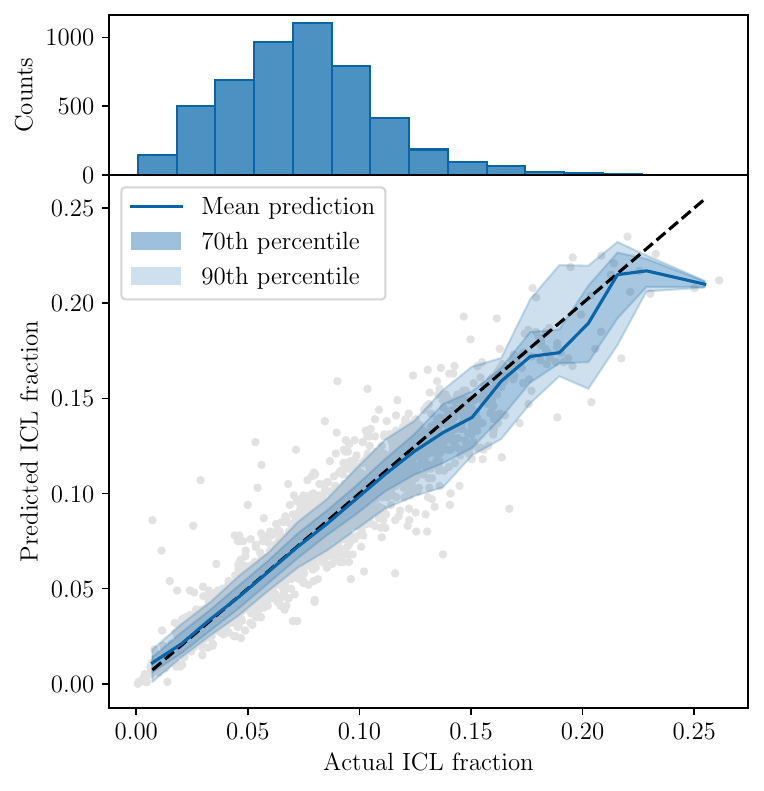}
    \caption{Top plot: Distribution of the measured ICL fractions in the validation data labels. Bottom plot: Performance of the model on the artificial dataset, plotting predicted values against actual values. Points have been binned into 20 equally spaced bins, with the solid line showing the median, and the solid colors marking regions within which 75\% (dark blue) and 90\% (lighter blue) of the points fall. The black dashed line shows the ideal one to one relationship.}
    \label{fig:performance_ad}
\end{figure}

Figure~\ref{fig:performance_ad} shows a plot of the model's final performance on the validation split (5,000 images) of the artificial dataset after the initial training was completed. The top plot shows the distribution of ICL fractions over the validation split. The bottom plot shows the ICL fraction value predicted by the model as a function of the actual ICL fraction value, with the values binned into 20 equally spaced bins for clarity, and the scatter in each bin represented by the filled regions. A perfect model would predict every value correctly, and would be represented as a perfect one to one relationship on the plot, which is marked as a black dashed line. We see that in general, the model predicts very close to the one to one relationship with only a small amount of scatter. The mean absolute error (MAE) between the model's predictions and the actual fraction is 0.00509. 

The mean predictions of the model as shown in Figure \ref{fig:performance_ad} appear to worsen at the higher range of ICL fractions, with the MAE in the second highest and highest bins being 0.0116 and 0.0447 respectively. However, given the small number of points in these bins (2 in each bin), it is not clear how representative these values are of the true ability of the model in this range. There are very few clusters in our sample that have an ICL fraction greater than 0.2 - only 129 clusters from our full artificial dataset are in this range (0.258\% of the dataset). Although, as seen in Figure \ref{fig:performance_ad}, the model does qualitatively appear to have been able to learn to distinguish these from lower ICL fraction images, it is clear from the MAE in the final bins that this range of fractions is less well characterized by the model than clusters with ICL fractions below 0.2 due to the lack of training samples. This could cause it to predict with more scatter, as well as preferentially predict slightly closer to the mean to minimize the loss penalty of mistakenly predicting a high ICL fraction. 

Despite these small inaccuracies, we can see that the model has certainly learned how to distinguish high ICL images from low ICL images for this dataset, and generally achieves a high accuracy in its predictions.

\subsection{Real data}
\begin{figure}
    \centering
    \includegraphics[width=0.95\linewidth]{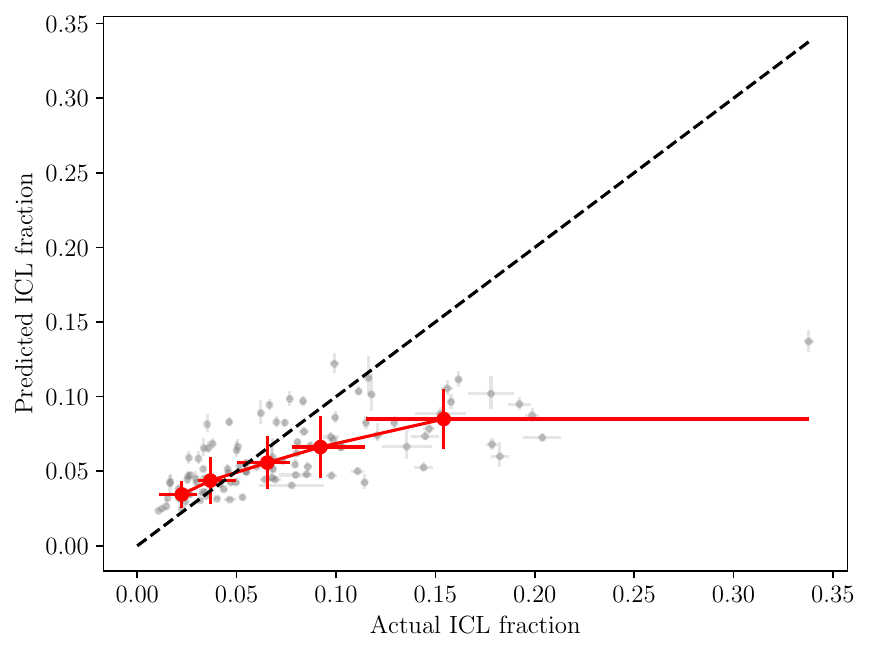}
    \caption{Performance of the model on the real dataset before fine-tuning, plotting predicted values against actual values. Grey points show all data points, where vertical error bars are the $1\sigma$ confidence intervals calculated from the output probability distributions, and horizontal error bars show the uncertainty in the manual measurements. Red points are median values in five bins with equal numbers of points, with horizontal error bars showing the extent of each bin.}
    \label{fig:performance_prefinetune}
\end{figure}

Figure~\ref{fig:performance_prefinetune} shows the model's performance on the real data, before any fine-tuning has taken place. We also bin the data into 5 bins with equal numbers of points to better see the general trend in the model's predictions. There is a clear linear upwards trend in the model's predictions, showing that the model has been able to apply its learning to this new data domain to an extent, but the trend is flatter than the ideal one to one relationship shown as the black dashed line. The model also appears to be overconfident in its predictions, shown by the extremely small vertical error bars, which represent the $1\sigma$ confidence intervals calculated from the output probability distributions. The MAE in the model's predictions is 0.0295. Fine-tuning on some amount of real data should help the model adjust to the differences in domain, and reevaluate its confidence in its predictions. 

\begin{figure}
    \centering
    \includegraphics[width=0.95\linewidth]{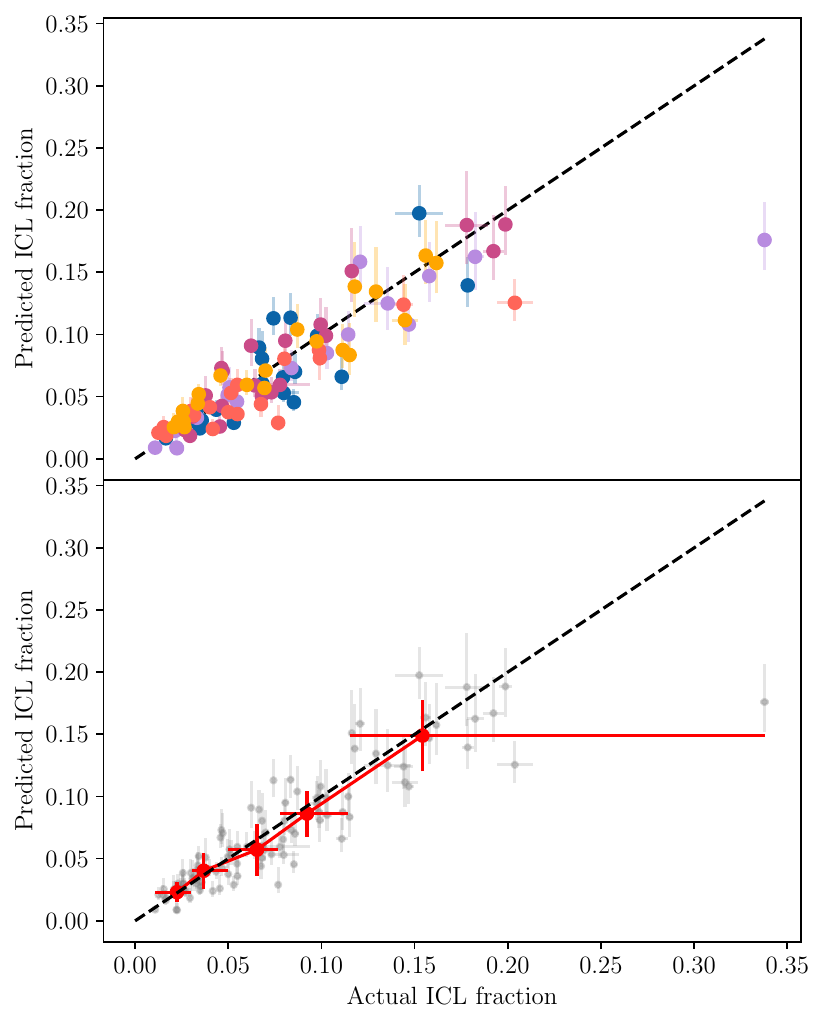}
    \caption{Performance of the model on the real dataset after fine-tuning, plotting predicted values against actual values. Top plot: Results from each fine-tuning run indicated by different colors, with different samples used for testing each time. Vertical error bars are the $1\sigma$ confidence intervals calculated from the output probability distributions, and horizontal error bars show the uncertainty in the manual measurements. Bottom plot: All data points are shown in grey. Red points showing median values in five bins with equal numbers of points, with horizontal error bars showing the extent of each bin.}
    \label{fig:performance_finetuned}
\end{figure}

Figure~\ref{fig:performance_finetuned} shows the model's performance after fine-tuning on the real dataset. We use an 80-20 split for training and testing, however, due to the small size of the real dataset, this equates to only 20 points used for testing. We therefore run the fine-tuning process on 5 different splits of the data, with different points used for testing each time, in order to collate the results and evaluate the effect of fine-tuning on the model's performance in a statistically significant way. Each model is trained from the same initial point, using weights saved from the initial training on the artificial dataset. The top panel in Figure~\ref{fig:performance_finetuned} shows which predictions come from which split in the data, whereas the bottom panel shows all the data points, binned into 5 bins, each with equal numbers of points. 

The top panel shows that the splits are sufficiently random to give the model a good range of data points each time, and also shows that the fine-tuning of the model is stable, since each split shows a similar testing performance despite the different partitions of training and testing data used. 

\begin{figure}
    \centering
    \includegraphics[width=0.95\linewidth]{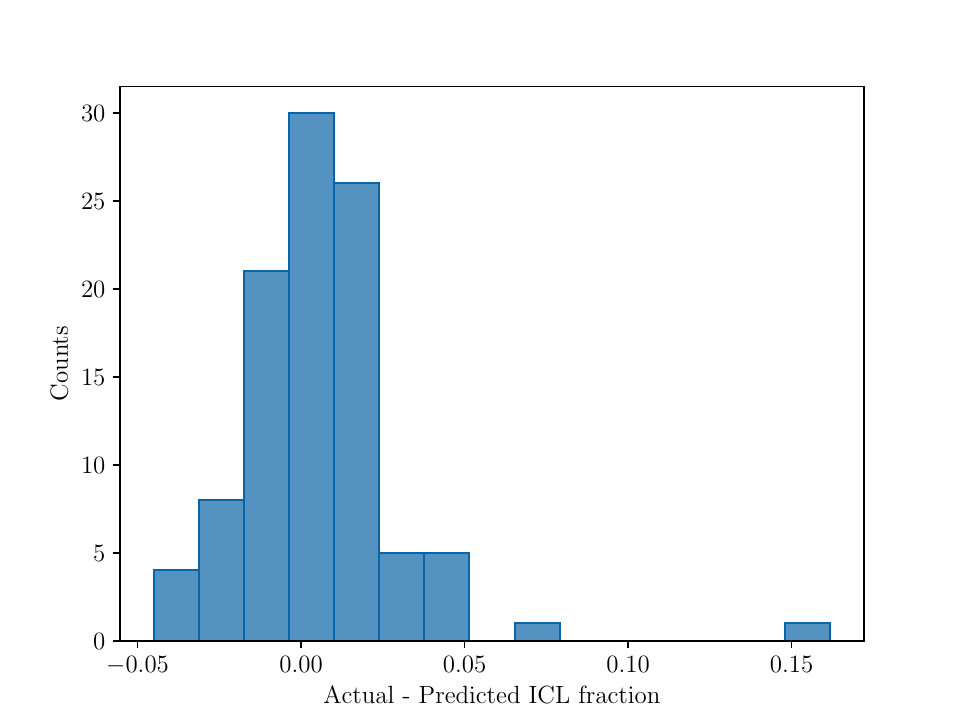}
    \caption{Scatter in ICL fraction predictions of the model with respect to actual fractions in the fine-tuning dataset. The distribution has a mean of -0.005 and rms of 0.025.}
    \label{fig:scatter}
\end{figure}

We can see from the binned trend in the bottom panel that fine-tuning has successfully corrected the bias at higher ICL fractions that we observed in the pre-fine-tuning performance in Figure~\ref{fig:performance_prefinetune}, with the binned points now falling very close to the one to one relationship. The MAE for the model now is 0.0159. The model has also reevaluated its confidence intervals after being trained on this new data. These now better reflect the actual scatter in the model's predictions. On average, the model now has excellent agreement with the fractions measured manually with the surface brightness method, however, we do observe some scatter in this relationship. In Figure~\ref{fig:scatter} we show the extent of this scatter for our dataset. Generally it is of the order of a few percent, with a mean of -0.005 and rms of 0.025, but it does reach a maximum of 0.162 in the worst case. 

This maximum is due to a significant outlier in the measured fraction, evident in Figure \ref{fig:performance_finetuned}, which has a measured ICL fraction of 0.338±0.003. This is 0.134 higher than the next highest fraction. This outlier poses an issue for the model, which underpredicts the measured fraction significantly (prediction $0.176^{+0.031}_{-0.024}$). One explanation for this is that the model characterizes this cluster poorly due to it being a very large outlier. When this cluster is in the test split during fine-tuning (where the prediction shown in Figure \ref{fig:performance_finetuned} comes from), the model will never see any cluster with a fraction this high, and is therefore not able to recognize it correctly during testing. In actual use of the model, this would be mitigated since this cluster would form part of the fine-tuning set. The other possibility is that some element of this specific image poses a challenge for the model. We investigated whether the presence of a significant foreground galaxy is causing the model to mistakenly include it as part of the cluster, resulting in the underestimated ICL fraction. However, masking the galaxy in question did not result in a significant difference in the model’s prediction (now predicting $0.196^{+0.032}_{-0.024}$, within the uncertainties of the original prediction). Without other examples of very high ICL fraction clusters, it is unclear whether this is an issue with this particular cluster, or a systematic issue with all very high ICL fraction clusters. In any case, we must always be cautious about trusting the particular prediction for one single cluster by the model, particularly for higher fractions, until the fine-tuning sample is sufficiently expanded to provide more training with high ICL fraction clusters.

\section{Discussion}\label{sec:discussion}
We have shown that our machine learning model is capable of accurately predicting ICL fractions in cluster images, where the provided images require no manual preprocessing (i.e. the images given to the model are exactly as downloaded from HSC-SSP, with automatic application of the bright star masks provided) and the model is not provided any other external information. Even with only a small amount of available data and measurements (just 101 real cluster images), through the use of transfer learning the model was able to effectively learn the task and adapt its learning from an artificial domain to the real domain. However, there are concerns that come with any machine learning model that can influence how useful these models are in real science applications. In this section, we discuss how our model may be affected by these issues. 

\subsection{Manual measurements}
A machine learning model is only as good as the data it is trained on, and without any real ``ground truth" ICL dataset available for training, and with the wide variety of different measurement methods available, it is important to consider the systematics involved in the method we chose to use for our data labeling. 

In this work we used the surface brightness cut method. This is one of the simplest methods to implement, making it relatively easy to automate for our large artificial dataset. It also has the advantage that it makes no assumptions about the morphology of either the BCG or the ICL. The clear drawback to this method is the relatively arbitrary choice of surface brightness threshold, and the fact that this method cannot measure the part of the ICL that falls over the BCG or any other galaxies.  

% \subsection{Comparison to other studies}

\citet{brough_preparing_2024} made a comparison of the performance of eight different ICL measurement methods applied to 61 mock images of galaxy clusters from simulations. They found that, within the uncertainties, the mean ICL fraction measured by the eight different methods on the same cluster images were consistent, ranging from $0.09\pm0.02$ to $0.17\pm0.08$, with a mean over all methods of $0.13\pm0.05$. The two surface brightness threshold methods found a mean ICL fraction of $0.14\pm0.04$ (`Martinez-Lombilla') and $0.14\pm0.03$ (`Montes'). Our measurement method, applied to our fine-tuning dataset, finds a mean ICL fraction of $0.075\pm0.055$, and for our artificial dataset the mean is $0.075\pm0.034$. These values are lower than the mean found by \citet{brough_preparing_2024}. However, this is not surprising when we consider the differences between the datasets. \cite{brough_preparing_2024} are measuring the ICL fraction in mock images from simulations rather than real observations, and make their measurements within a 1 Mpc aperture, significantly larger than ours. They also simulated a limiting surface brightness of $\mu_r=30.3$ mag/arcsec$^2$ to match the estimated LSST 10 year limiting surface brightness. This limit is deeper than our images, which reach a maximum of $\mu_r = 29.8$ mag/arcsec$^2$ for the Ultradeep layer of HSC \citep{martinez-lombilla_galaxy_2023}.

The only other work measuring the ICL fraction in HSC-SSP images is \citet{furnell_growth_2021}. They measured the ICL in 18 clusters detected from the XMM Cluster Survey data that fall in the deep HSC-SSP field. Five of these clusters also appear in our final fine-tuning sample. They also use the surface brightness cut method, however, their method deviates in other ways. Most significantly, they use a different surface brightness threshold to perform their measurement ($\mu_B = 25$ mag/arcsec$^2$). Assuming an age of 2 Gyr and [Fe/H] of -0.4 for the ICL (e.g. \citealp{montes_intracluster_2018}) giving a color of $B-r=0.74$ \citep{vazdekis_uv-extended_2016}, this corresponds to a surface brightness threshold of 24.26 mag/arcsec$^2$ in the $r$ band, significantly brighter than our threshold of 26 mag/arcsec$^2$. A brighter threshold will lead to more light from the brighter parts of the cluster being included in the ICL, naturally leading to higher fractions. They also make their measurements within varying radii, depending on the $R_{500}$ of their clusters as estimated from X-ray temperatures, with a mean over all clusters of $\sim600$ kpc, on average significantly larger than our chosen radius of 300 kpc. As expected, \citet{furnell_growth_2021} find a mean fraction of $0.243\pm0.002$, much higher than ours. The differences in our measurement methods make it difficult to draw conclusions from these comparisons, despite having some of the same clusters in our samples. 

\citet{martinez-lombilla_galaxy_2023} also use HSC-SSP images, although they measured the diffuse light within a galaxy group (intragroup light, or IGL), rather than a cluster. They performed their measurements in $g$, $r$, and $i$ bands, and used both a composite model method and a surface brightness threshold method at two different thresholds. The method most comparable to ours (surface brightness threshold of 26 mag/arcsec$^2$ in the $r$ band, within a radius of $\sim275$ kpc), found an IGL fraction of $0.086\pm0.024$, which agrees well with the mean ICL fraction that we find in our sample. However, the IGL fraction that they find using other methods and other bands varies considerably in their studied group, from $0.016\pm0.043$ to $0.365\pm0.022$. This again demonstrates the importance of building a homogeneous sample of ICL measurements using a consistent measurement method, as even when using the same data, differences in measurement method can make it very difficult to compare between studies.

\begin{figure}
    \centering
    \includegraphics[width=0.95\linewidth]{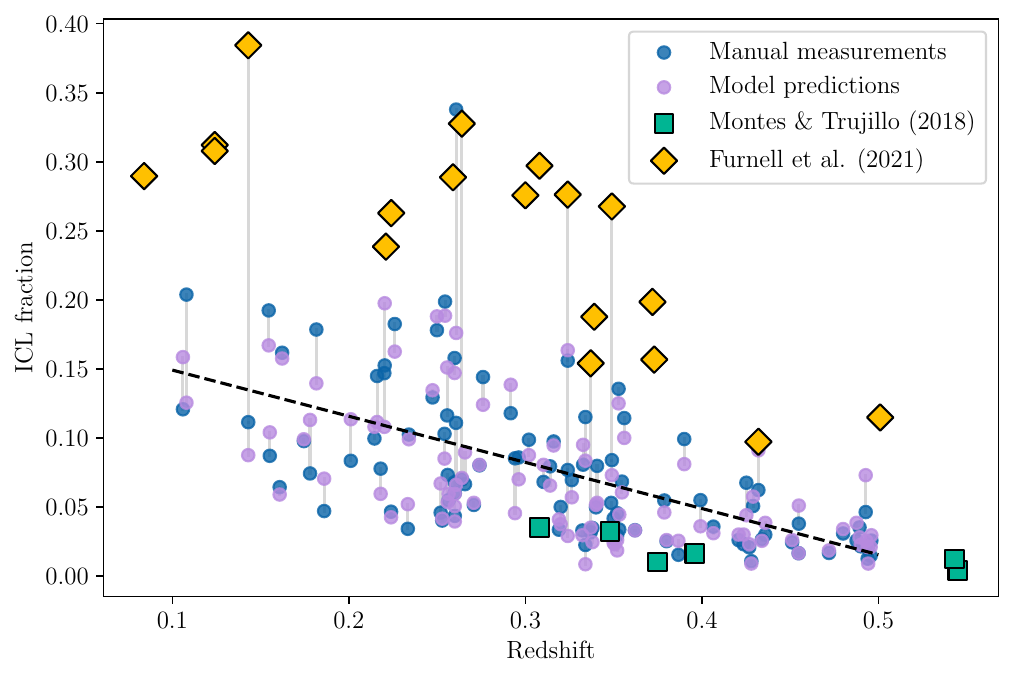}
    \caption{ICL fraction as a function of redshift. Blue and purple circles represent our manually measured and model predicted fractions, respectively, and the black dashed line shows the line of best fit for our manual measurements. Teal squares are the fractions from \citet{montes_intracluster_2018}, and yellow diamonds are the fractions from \citet{furnell_growth_2021}. Grey lines connect fractions that correspond to the same cluster.} 
    \label{fig:redshift}
\end{figure}

We can also consider any dependence on redshift that we observe in our sample. In Figure~\ref{fig:redshift} we present the fractions we measure for our clusters, as well as the fractions predicted by the model. We also plot measurements from \citet{furnell_growth_2021}, and \citet{montes_intracluster_2018}, as their clusters are within a similar redshift range to ours. Grey lines connect different measurements for the same cluster. 

We find a negative relationship between our measured ICL fractions and redshift. The line of best fit for the manual measurements is shown in Figure~\ref{fig:redshift} by the black dashed line and is given by $f_{\textrm{ICL}} = (-0.33 \pm 0.04)z + (0.18 \pm 0.01)$. The relationship given by the model predictions is $f_{\textrm{ICL}} = (-0.30 \pm 0.04)z + (0.17 \pm 0.01)$. This relationship is consistent within uncertainties between the manual measurements and the model predictions, showing that the model does not show a particular redshift-dependent bias in its performance. The strength of this negative relationship is similar to that found by \citet{montes_intracluster_2018}, although our fractions are on average slightly higher than theirs. They also use a surface brightness threshold method, using a threshold of $\mu_V = 26$ mag/arcsec$^2$ and images that are 600 kpc on each side. \citet{furnell_growth_2021} measure fractions much higher than ours on average, and also find a stronger correlation with redshift. As discussed previously, this is likely due to several important differences between our method and theirs.

There are a number of assumptions that we made in our measurements, particularly of the artificial dataset, that could potentially affect our model's accuracy. We noted that our use of smooth ICL profiles in the dataset generation could impact our model's ability to identify ICL with more asymmetric morphologies, however we do not find this to be a particular issue after the fine-tuning of the model on real data that does include examples of more asymmetric ICL profiles. In our artificial dataset measurements, we were not able to use cluster membership information, and instead assumed that all galaxies in the cutout belonged to the cluster. This would generally lead to an overestimate of cluster flux and therefore could lead to underestimates in ICL fraction, and we did in fact observe a tendency of the model to underestimate ICL fraction before fine-tuning, as shown in Figure \ref{fig:performance_prefinetune}. However, fine-tuning appears to have been able to correct this issue as shown in Figure \ref{fig:performance_finetuned}. Our automatic background subtraction method as used for the artificial dataset was not always perfect, and we did observe some over or under-subtraction that required adjustment in the manual measurements. However, the tests we present in Section \ref{sec:method} and the model's ability to easily learn to match the manual measurements with more accurate background subtraction shows that it is not overly sensitive to the exact background subtraction method. We do not find these potential issues with the data generation and automatic measurements to have lasting effects on the model's accuracy as shown by its higher performance after the fine-tuning stage. 

There are other assumptions made while measuring the fine-tuning dataset that could affect the uncertainties in our manual measurements. Although we do include cluster membership information in our manual measurements, these are based on cuts in photometric redshift, a common way to estimate cluster members in the absence of spectroscopic information (e.g. \citealp{burke_coevolution_2015}; \citealp{montes_intracluster_2018}; \citealp{furnell_growth_2021}), but which could lead to misidentifications. 

Our background subtraction is also a source of measurement uncertainty, even though in the case of the manual measurements we visually inspect the result for each cluster. We calculate our measurement uncertainty based on the sensitivity of the ICL fraction to background subtraction parameter choice, and find that this aspect should not influence our measurements significantly. 

Downsizing the images to 224$\times$224 pixels for the model input may also be impacting its predictions as compared to our measurements, as images that are lower redshift are more heavily binned due to their larger original size than images at higher redshifts. However, \cite{brough_preparing_2024} found that binning images by different amounts before measurement did not have a significant impact on the ICL fractions found, and we do not observe sensitivity in the model to this as shown by the lack of dependence of the model's performance on redshift in Figure \ref{fig:redshift}. 

Finally, our choice of radius is another important assumption in our method. We chose a radius of 300 kpc, owing to the fact that this generally corresponds to the ICL detection limit (e.g. \citealp{demaio_origin_2015}; \citealp{jimenez-teja_unveiling_2018}; \citealp{montes_buildup_2021}) and the fact that we cannot estimate physical cluster sizes for all of our clusters with our available data. However, it is possible that using a fixed physical radius will have an impact on our measured values, particularly at low redshifts where the ICL is more extended on the sky. To investigate the impact of our chosen radius limit on our measurements, we manually measure ICL fractions for a sample of clusters out to the $R_{500}$ scale radius (the radius containing a volume with a mean density 500 times the critical density at that redshift) and compare the fractions to our measurements within 300 kpc. To estimate $R_{500}$ we crossmatch our Deep/Ultradeep CAMIRA clusters with the Galaxy and Mass Assembly (GAMA) Galaxy Group Catalogue \citep{robotham_galaxy_2011} and find 13 matches with $\log(M_{200})$ ranging from 13.7 $M_\odot$ to 15.0 $M_\odot$. We use a scaling relation to calculate $R_{500}$ from the measured velocity dispersions of the GAMA groups \citep{zhang_hiflugcs_2011} Over this subsample, the mean $R_{500}$ is 640 ± 220 kpc.

We find that the ICL fractions measured within $R_{500}$ increase by a mean of 0.026 and up to a maximum of 0.073. Clusters with larger $R_{500}$ have larger differences in the measured fraction compared to the measurements taken within a radius of 300 kpc, however the differences appear to be independent of cluster redshift. This indicates that there is some ICL that is missed when measuring within a radius of 300 kpc as compared to $R_{500}$. However, we note that there is currently no consensus on the best radius to use for measuring the ICL fraction, and there are a number of approaches taken in the literature ranging from a consistent physical radius, to cluster-dependent radius measurements. We have chosen to use a physical radius of 300 kpc as it is easy to maintain consistency amongst a photometrically selected cluster sample where scale radius estimates are not available and balances with the machine learning requirements, where significantly larger radii (e.g. 1 Mpc) pose issues with the spatial resolution of pixels after they have been resized to the 224 x 224 pixels necessary for the model input. However, we note that our choice of a 300 kpc radius is a free parameter that could complicate comparison with measurements that are made with a different radius. Our focus is to enable the construction of a large homogeneous sample, so it is more important to ensure that the choice of radius can be easily applied to all of our cluster measurements. 

There are significant difficulties with comparing manual ICL fraction measurements between different samples and different methods, as anything from differing surface brightness limits to choice in observed wavelength, surface brightness threshold, or radius can have a significant impact on the measured fraction. Our fractions are consistent with previous comparable studies, and although some assumptions we have made introduce some uncertainties that may be adopted by the model, we do not find these to contribute significantly to uncertainties in the model's predictions. 

\subsection{Interpretability of machine learning outputs}
An issue in the application of machine learning models is the trustworthiness of the model's outputs. Models are often treated as ``black boxes", meaning that the user sometimes has no idea why the model has produced a particular answer for a particular input.

In particular for regression models, many models have no way of indicating confidence in their answer, meaning that there is no way to distinguish between trustworthy and untrustworthy outputs. We address this issue by using a probabilistic output. In the model's training, it aims to maximize the likelihood of the correct answer, and is penalized based on how unlikely the correct answer was under the predicted distribution. This means that if the model is unsure, it is incentivized to predict a more flat probability distribution to minimize the penalty. Since we retain this probability distribution, we therefore have information about how confident the model is in its prediction. This is demonstrated in the error bars given in Figures \ref{fig:performance_prefinetune} and \ref{fig:performance_finetuned}.

Another concern about machine learning models is their ability to find and exploit shortcuts in data that they have been given, so although it can appear from the test results that the model has learned the task, in fact it may be exploiting peculiarities in the dataset to ``cheat" the answer and not performing the task as intended. This is a well-known problem across many domains of machine learning, sometimes known as shortcut learning (see \citealp{geirhos_shortcut_2020} for a review). Given that our model is easily able to transfer its learning from one dataset to another (the artificial dataset to the real images), this is unlikely to be the case, and indicates that the model has in fact learned the real task of identifying the ICL and not overfit to some underlying pattern in the artificial dataset. However, to check this assumption, we can use saliency maps, which highlight pixels in input images that the model deems to be important when making its decision. These can be very useful to check that a model is not exploiting shortcuts in the data (e.g. \citealp{zech_variable_2018}).

The method we use to generate these saliency maps is GradCAM \citep{selvaraju_grad-cam_2020}. GradCAM uses the magnitude of gradients flowing into the last convolutional layer in the network (the last layer to retain spatial information from the image) to assign importance to different features in the network. An activation map can then be generated over the original image that combines all feature map activations weighted by importance. This can be used to check that the model is focusing on areas of the image that make sense. 

\begin{figure}
    \centering
    \includegraphics[width=0.95\linewidth]{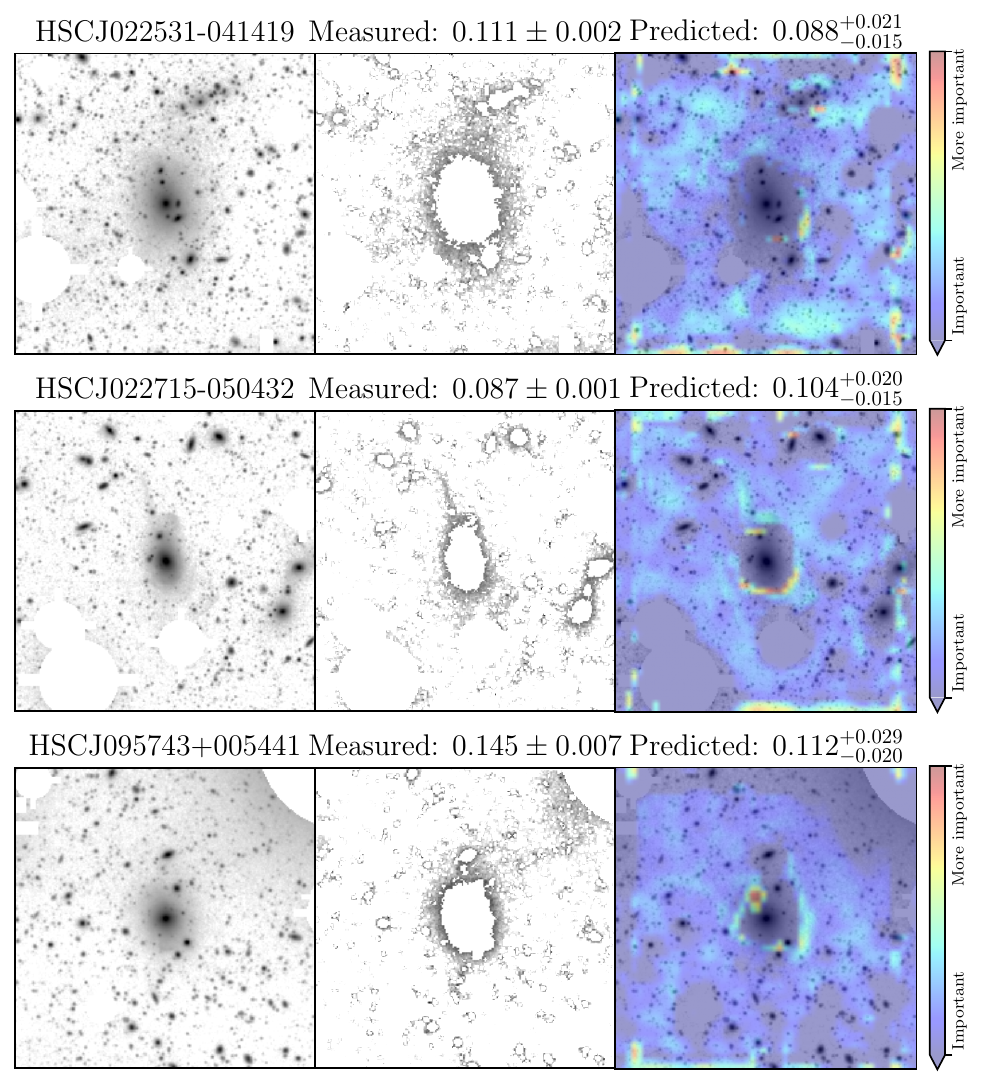}
    \caption{GradCAM outputs on three example cluster images from the fine-tuning dataset. Top row: an example of a typical activation map. Middle row: an example of the model activating on a tidal tail and shell. Bottom row: an example of the model ignoring a background gradient due to a poorly masked star. Left column: Original images from the fine-tuning dataset. Middle column: Input images, masked and thresholded to bring out the low surface brightness light. Right column: Activation maps overlaid on the original images, where red indicates stronger activation. }
    \label{fig:gradcam}
\end{figure}

Figure~\ref{fig:gradcam} shows three GradCAM outputs for images from the fine-tuning dataset, where we ask it to highlight areas of the image that are important for predicting a high ICL fraction. Note that this is not the same as asking the model to highlight areas where ICL exists, rather, the model shows us what areas in the image it is using to distinguish between it having a low ICL fraction and it having a high ICL fraction. For the sake of comparison, we also show in Figure~\ref{fig:gradcam} a thresholded version of each input image to show where the low surface brightness light is actually located in the image. A shortcoming of GradCAM is that it does not tell us exactly how the model is using these areas in making its decision, only the relative importance. This means that as a tool, it is most suited for checking that the model is using areas of the image that indicate it is performing the task as intended. 

In Figure~\ref{fig:gradcam} we see that the model is focusing primarily on the low surface brightness areas of the image while largely ignoring the bright, central parts of the galaxy, showing that it is considering relevant parts of the image to make an informed conclusion. Its attention to these areas correctly applied to each image makes it unlikely to be exploiting peculiarities in the dataset. There are some points of interest in these activation maps. Firstly, we see that there is a tendency to activate along the borders of the images. It is possible that these indicate the model checking for background gradients to ignore -- in fact, we see in the example in the bottom row that the model is capable of disregarding the background gradient that is present in that image due to a large star that is not fully masked in the top corner. There is also the possibility that this is an artifact due to the model's architecture. \citet{alsallakh_mind_2020} found similar artifacts in their images to be connected to padding in the model's convolutional layers, which has also been connected to a model's ability to learn spatial information \citep{semih_kayhan_translation_2020}. With only these saliency maps, the exact reason for the activation along the borders remains unclear. We also see that, interestingly, the model shows interest in tidal features -- in the middle example, the central galaxy has a clear tidal tail and shells, which correspond to bright spots in the saliency maps. 

This type of visual interpretability is still quite new in machine learning, and as the tools improve we will be able to have greater insight into how models make their decisions. GradCAM allows us to see that the model is not exploiting shortcuts in the dataset to ``cheat" the task, and really has learned some underlying rules in the images to using the low surface brightness light to predict the ICL fraction. This indicates that its high performance on the test data can be trusted.

\section{Conclusion}
The problem of systematics introduced by choice of measurement method is still an unsolved one in the field of ICL measurements, which is partly what motivates the construction of a large, homogeneous sample with consistent assumptions and systematic uncertainties. Here we present a model trained using one reliable and commonly used measurement method, showing that machine learning models are capable of learning to perform a task on par with manual measurement. As understanding of different ICL measurement methods evolves, the data that the model is trained on can be developed further. In the future, it will be possible to fine-tune the model on larger samples made with different methods or even combinations of methods, in order to compare the quality of the predictions made.

The main goal of developing a machine learning method such as this is to enable fast processing of vast amounts of data that more traditional measurement methods are unable to cope with, which will be available in new surveys such as LSST as well as Euclid (e.g. \citealp{kluge_euclid_2024}). From our results, we can see that our model is capable of replicating measurements made with traditional methods without any of the manual inspection and preprocessing that would normally be necessary. 

A distinct advantage to the throughput of the model is its ability to process a batch of input images at once, in parallel. The size of the batch that can be processed at once is limited by the available memory. On the single GPU that we use for training the model, which has 31GB of memory, our model is able to process a maximum of 500 samples at once in a matter of seconds. 

Of course, the model is not always perfect, and is not always able to perfectly replicate the manual measurements. This scatter is relatively small, but given that the model in itself is not physically motivated, for more detailed study and measurement of individual clusters, we will still need traditional measurement methods to get a complete picture of the ICL. More detailed measurements using traditional methods will also provide the better understanding of the systematics and assumptions involved in the different methods needed to advance this field. With this in place we will also be able to continue to train this model to make its predictions more robust and accurate. Currently, the model is trained on just 101 real cluster images and measurements, and already achieves high performance, so even a small amount of extra data could help to improve its predictions. For now, the model's best use is in quickly and automatically processing a very large number of image samples in order to study large-scale trends in the ICL fraction, and identify any particularly interesting, outlying systems for detailed follow up studies.

In this work, we have shown that a machine learning model can accurately predict the ICL fraction in cluster images given only a small number of real training images by leveraging transfer learning. Its ability to quickly process many images at once makes it a good candidate for dealing with the massive amounts of data that large imaging surveys such as LSST and Euclid will provide us. More data and development of the various manual measurement methods used in the ICL field will be used to refine the fine-tuning of the model and continue to improve its performance. All of the code used for creating, training, and testing the model is made available to anyone interested in using it on GitHub at \url{https://github.com/lpcan/MICL} \citep{canepa_lpcanmicl_2024}. 

%% IMPORTANT! The old "\acknowledgment" command has be depreciated. It was
%% not robust enough to handle our new dual anonymous review requirements and
%% thus been replaced with the acknowledgment environment. If you try to 
%% compile with \acknowledgment you will get an error print to the screen
%% and in the compiled pdf.
%% 
%% Also note that the akcnowlodgment environment does not support long amounts of text. If you have a lot of people and institutions to acknowledge, do not use this command. Instead, create a new \section{Acknowledgments}.
\section*{Acknowledgements}
We thank the anonymous referee for their thoughtful report that has improved the paper.

SB acknowledges funding support from the Australian Research Council through a Discovery Project DP190101943. MM acknowledges support from the project RYC2022-036949-I financed by the MICIU/AEI/10.13039/501100011033 and by FSE+. Parts of this research were supported by the Australian Research Council Centre of Excellence for All Sky Astrophysics in 3 Dimensions (ASTRO 3D), through project number CE170100013. Parts of this research were supported by the Astronomical Society of Australia (ASA), through the Student Travel Assistance Scheme (STAS).

This research includes computations using the computational cluster Katana supported by Research Technology Services at UNSW Sydney.

The Hyper Suprime-Cam (HSC) collaboration includes the astronomical communities of Japan and Taiwan, and Princeton University. The HSC instrumentation and software were developed by the National Astronomical Observatory of Japan (NAOJ), the Kavli Institute for the Physics and Mathematics of the Universe (Kavli IPMU), the University of Tokyo, the High Energy Accelerator Research Organization (KEK), the Academia Sinica Institute for Astronomy and Astrophysics in Taiwan (ASIAA), and Princeton University. Funding was contributed by the FIRST program from the Japanese Cabinet Office, the Ministry of Education, Culture, Sports, Science and Technology (MEXT), the Japan Society for the Promotion of Science (JSPS), Japan Science and Technology Agency (JST), the Toray Science Foundation, NAOJ, Kavli IPMU, KEK, ASIAA, and Princeton University. 

This paper makes use of software developed for Vera C. Rubin Observatory. We thank the Rubin Observatory for making their code available as free software at http://pipelines.lsst.io/.

This paper is based on data collected at the Subaru Telescope and retrieved from the HSC data archive system, which is operated by the Subaru Telescope and Astronomy Data Center (ADC) at NAOJ. Data analysis was in part carried out with the cooperation of Center for Computational Astrophysics (CfCA), NAOJ. We are honored and grateful for the opportunity of observing the Universe from Maunakea, which has the cultural, historical and natural significance in Hawaii.

\bibliography{bibliography}{}

\begin{thebibliography}{}
\expandafter\ifx\csname natexlab\endcsname\relax\def\natexlab#1{#1}\fi
\providecommand{\url}[1]{\href{#1}{#1}}
\providecommand{\dodoi}[1]{doi:~\href{http://doi.org/#1}{\nolinkurl{#1}}}
\providecommand{\doeprint}[1]{\href{http://ascl.net/#1}{\nolinkurl{http://ascl.net/#1}}}
\providecommand{\doarXiv}[1]{\href{https://arxiv.org/abs/#1}{\nolinkurl{https://arxiv.org/abs/#1}}}

\bibitem[{Adami {et~al.}(2013)Adami, Durret, Guennou, \& Rocha}]{adami_diffuse_2013}
Adami, C., Durret, F., Guennou, L., \& Rocha, C.~D. 2013, Astronomy \& Astrophysics, 551, A20, \dodoi{10.1051/0004-6361/201220282}

\bibitem[{Aihara {et~al.}(2019)Aihara, AlSayyad, Ando, Armstrong, Bosch, Egami, Furusawa, Furusawa, Goulding, Harikane, Hikage, Ho, Hsieh, Huang, Ikeda, Imanishi, Ito, Iwata, Jaelani, Kakuma, Kawana, Kikuta, Kobayashi, Koike, Komiyama, Li, Liang, Lin, Luo, Lupton, Lust, MacArthur, Matsuoka, Mineo, Miyatake, Miyazaki, More, Murata, Namiki, Nishizawa, Oguri, Okabe, Okamoto, Okura, Ono, Onodera, Onoue, Osato, Ouchi, Shibuya, Strauss, Sugiyama, Suto, Takada, Takagi, Takata, Takita, Tanaka, Terai, Toba, Uchiyama, Utsumi, Wang, Wang, \& Yamada}]{aihara_second_2019}
Aihara, H., AlSayyad, Y., Ando, M., {et~al.} 2019, Publications of the Astronomical Society of Japan, 71, 114, \dodoi{10.1093/pasj/psz103}

\bibitem[{Aihara {et~al.}(2022)Aihara, AlSayyad, Ando, Armstrong, Bosch, Egami, Furusawa, Furusawa, Harasawa, Harikane, Hsieh, Ikeda, Ito, Iwata, Kodama, Koike, Kokubo, Komiyama, Li, Liang, Lin, Lupton, Lust, MacArthur, Mawatari, Mineo, Miyatake, Miyazaki, More, Morishima, Murayama, Nakajima, Nakata, Nishizawa, Oguri, Okabe, Okura, Ono, Osato, Ouchi, Pan, Plazas~Malagón, Price, Reed, Rykoff, Shibuya, Simunovic, Strauss, Sugimori, Suto, Suzuki, Takada, Takagi, Takata, Takita, Tanaka, Tang, Taranu, Terai, Toba, Turner, Uchiyama, Vijarnwannaluk, Waters, Yamada, Yamamoto, \& Yamashita}]{aihara_third_2022}
---. 2022, Publications of the Astronomical Society of Japan, 74, 247, \dodoi{10.1093/pasj/psab122}

\bibitem[{Alsallakh {et~al.}(2020)Alsallakh, Kokhlikyan, Miglani, Yuan, \& Reblitz-Richardson}]{alsallakh_mind_2020}
Alsallakh, B., Kokhlikyan, N., Miglani, V., Yuan, J., \& Reblitz-Richardson, O. 2020, arXiv preprint arXiv:2010.02178

\bibitem[{{Astropy Collaboration} {et~al.}(2022){Astropy Collaboration}, Price-Whelan, Lim, Earl, Starkman, Bradley, Shupe, Patil, Corrales, Brasseur, Nöthe, Donath, Tollerud, Morris, Ginsburg, Vaher, Weaver, Tocknell, Jamieson, van Kerkwijk, Robitaille, Merry, Bachetti, Günther, Aldcroft, Alvarado-Montes, Archibald, Bódi, Bapat, Barentsen, Bazán, Biswas, Boquien, Burke, Cara, Cara, Conroy, Conseil, Craig, Cross, Cruz, D'Eugenio, Dencheva, Devillepoix, Dietrich, Eigenbrot, Erben, Ferreira, Foreman-Mackey, Fox, Freij, Garg, Geda, Glattly, Gondhalekar, Gordon, Grant, Greenfield, Groener, Guest, Gurovich, Handberg, Hart, Hatfield-Dodds, Homeier, Hosseinzadeh, Jenness, Jones, Joseph, Kalmbach, Karamehmetoglu, Kałuszyński, Kelley, Kern, Kerzendorf, Koch, Kulumani, Lee, Ly, Ma, MacBride, Maljaars, Muna, Murphy, Norman, O'Steen, Oman, Pacifici, Pascual, Pascual-Granado, Patil, Perren, Pickering, Rastogi, Roulston, Ryan, Rykoff, Sabater, Sakurikar, Salgado, Sanghi, Saunders, Savchenko, Schwardt, Seifert-Eckert,
  Shih, Jain, Shukla, Sick, Simpson, Singanamalla, Singer, Singhal, Sinha, Sipőcz, Spitler, Stansby, Streicher, Šumak, Swinbank, Taranu, Tewary, Tremblay, de~Val-Borro, Van~Kooten, Vasović, Verma, de~Miranda~Cardoso, Williams, Wilson, Winkel, Wood-Vasey, Xue, Yoachim, Zhang, Zonca, \& {Astropy Project Contributors}}]{astropy_collaboration_astropy_2022}
{Astropy Collaboration}, Price-Whelan, A.~M., Lim, P.~L., {et~al.} 2022, The Astrophysical Journal, 935, 167, \dodoi{10.3847/1538-4357/ac7c74}

\bibitem[{Barfety {et~al.}(2022)Barfety, Valin, Webb, Yun, Shipley, Boone, Hayden, Hlavacek-Larrondo, Muzzin, Noble, Perlmutter, Rhea, Wilson, \& Yee}]{barfety_assessment_2022}
Barfety, C., Valin, F.-A., Webb, T. M.~A., {et~al.} 2022, The Astrophysical Journal, 930, 25, \dodoi{10.3847/1538-4357/ac61dd}

\bibitem[{Bosch {et~al.}(2018)Bosch, Armstrong, Bickerton, Furusawa, Ikeda, Koike, Lupton, Mineo, Price, Takata, Tanaka, Yasuda, AlSayyad, Becker, Coulton, Coupon, Garmilla, Huang, Krughoff, Lang, Leauthaud, Lim, Lust, MacArthur, Mandelbaum, Miyatake, Miyazaki, Murata, More, Okura, Owen, Swinbank, Strauss, Yamada, \& Yamanoi}]{bosch_hyper_2018}
Bosch, J., Armstrong, R., Bickerton, S., {et~al.} 2018, Publications of the Astronomical Society of Japan, 70, S5, \dodoi{10.1093/pasj/psx080}

\bibitem[{Bradley {et~al.}(2023)Bradley, Sipőcz, Robitaille, Tollerud, Vinícius, Deil, Barbary, Wilson, Busko, Donath, Günther, Cara, Lim, Meßlinger, Conseil, Burnett, Bostroem, Droettboom, Bray, Bratholm, Jamieson, Ginsburg, Barentsen, Craig, Morris, Perrin, Rathi, Pascual, Perren, \& Georgiev}]{bradley_astropyphotutils_2023}
Bradley, L., Sipőcz, B., Robitaille, T., {et~al.} 2023, astropy/photutils: 1.10.0,  Zenodo, \dodoi{10.5281/zenodo.1035865}

\bibitem[{Brough {et~al.}(2007)Brough, Proctor, Forbes, Couch, Collins, Burke, \& Mann}]{brough_spatially_2007}
Brough, S., Proctor, R., Forbes, D.~A., {et~al.} 2007, Monthly Notices of the Royal Astronomical Society, 378, 1507, \dodoi{10.1111/j.1365-2966.2007.11900.x}

\bibitem[{Brough {et~al.}(2020)Brough, Collins, Demarco, Ferguson, Galaz, Holwerda, Martinez-Lombilla, Mihos, \& Montes}]{brough_vera_2020}
Brough, S., Collins, C., Demarco, R., {et~al.} 2020, The {Vera} {Rubin} {Observatory} {Legacy} {Survey} of {Space} and {Time} and the {Low} {Surface} {Brightness} {Universe},  arXiv.
\newblock \url{http://arxiv.org/abs/2001.11067}

\bibitem[{Brough {et~al.}(2024)Brough, Ahad, Bahé, Ellien, Gonzalez, Jiménez-Teja, Kimmig, Martin, Martínez-Lombilla, Montes, Pillepich, Ragusa, Remus, Collins, Knapen, \& Mihos}]{brough_preparing_2024}
Brough, S., Ahad, S.~L., Bahé, Y.~M., {et~al.} 2024, Monthly Notices of the Royal Astronomical Society, 528, 771, \dodoi{10.1093/mnras/stad3810}

\bibitem[{Burke {et~al.}(2015)Burke, Hilton, \& Collins}]{burke_coevolution_2015}
Burke, C., Hilton, M., \& Collins, C. 2015, Monthly Notices of the Royal Astronomical Society, 449, 2353, \dodoi{10.1093/mnras/stv450}

\bibitem[{Canepa {et~al.}(2024)Canepa, Brough, Lanusse, Montes, \& Hatch}]{canepa_lpcanmicl_2024}
Canepa, L., Brough, S., Lanusse, F., Montes, M., \& Hatch, N. 2024, lpcan/{MICL}: {Paper} version,  Zenodo, \dodoi{10.5281/ZENODO.14376238}

\bibitem[{Cañas {et~al.}(2020)Cañas, Lagos, Elahi, Power, Welker, Dubois, \& Pichon}]{canas_stellar_2020}
Cañas, R., Lagos, C. d.~P., Elahi, P.~J., {et~al.} 2020, Monthly Notices of the Royal Astronomical Society, 494, 4314, \dodoi{10.1093/mnras/staa1027}

\bibitem[{Chilingarian {et~al.}(2010)Chilingarian, Melchior, \& Zolotukhin}]{chilingarian_analytical_2010}
Chilingarian, I.~V., Melchior, A.-L., \& Zolotukhin, I.~Y. 2010, Monthly Notices of the Royal Astronomical Society, 405, 1409, \dodoi{10.1111/j.1365-2966.2010.16506.x}

\bibitem[{Contini(2021)}]{contini_origin_2021}
Contini, E. 2021, Galaxies, 9, 60, \dodoi{10.3390/galaxies9030060}

\bibitem[{Contini {et~al.}(2014)Contini, De~Lucia, Villalobos, \& Borgani}]{contini_formation_2014}
Contini, E., De~Lucia, G., Villalobos, {\'A}., \& Borgani, S. 2014, Monthly Notices of the Royal Astronomical Society, 437, 3787, \dodoi{10.1093/mnras/stt2174}

\bibitem[{Contini {et~al.}(2023)Contini, Jeon, Rhee, Han, \& Yi}]{contini_intracluster_2023}
Contini, E., Jeon, S., Rhee, J., Han, S., \& Yi, S.~K. 2023, The Astrophysical Journal, 958, 72, \dodoi{10.3847/1538-4357/acfd25}

\bibitem[{Contreras-Santos {et~al.}(2024)Contreras-Santos, Knebe, Cui, Alonso~Asensio, Dalla~Vecchia, Cañas, Haggar, Mostoghiu~Paun, Pearce, \& Rasia}]{contreras-santos_characterising_2024}
Contreras-Santos, A., Knebe, A., Cui, W., {et~al.} 2024, Astronomy and Astrophysics, 683, A59, \dodoi{10.1051/0004-6361/202348474}

\bibitem[{Da~Rocha \& De~Oliveira(2005)}]{da_rocha_intragroup_2005}
Da~Rocha, C., \& De~Oliveira, C.~M. 2005, Monthly Notices of the Royal Astronomical Society, 364, 1069, \dodoi{10.1111/j.1365-2966.2005.09641.x}

\bibitem[{Da~Rocha {et~al.}(2008)Da~Rocha, Ziegler, \& Mendes~de Oliveira}]{da_rocha_intragroup_2008}
Da~Rocha, C., Ziegler, B.~L., \& Mendes~de Oliveira, C. 2008, Monthly Notices of the Royal Astronomical Society, 388, 1433, \dodoi{10.1111/j.1365-2966.2008.13500.x}

\bibitem[{Dalal {et~al.}(2021)Dalal, Strauss, Sunayama, Oguri, Lin, Huang, Park, \& Takada}]{dalal_brightest_2021}
Dalal, R., Strauss, M.~A., Sunayama, T., {et~al.} 2021, Monthly Notices of the Royal Astronomical Society, 507, 4016, \dodoi{10.1093/mnras/stab2363}

\bibitem[{DeMaio {et~al.}(2015)DeMaio, Gonzalez, Zabludoff, Zaritsky, \& Bradač}]{demaio_origin_2015}
DeMaio, T., Gonzalez, A.~H., Zabludoff, A., Zaritsky, D., \& Bradač, M. 2015, Monthly Notices of the Royal Astronomical Society, 448, 1162, \dodoi{10.1093/mnras/stv033}

\bibitem[{Desmons {et~al.}(2024)Desmons, Brough, \& Lanusse}]{desmons_detecting_2024}
Desmons, A., Brough, S., \& Lanusse, F. 2024, Monthly Notices of the Royal Astronomical Society, 531, 4070, \dodoi{10.1093/mnras/stae1402}

\bibitem[{Dillon {et~al.}(2017)Dillon, Langmore, Tran, Brevdo, Vasudevan, Moore, Patton, Alemi, Hoffman, \& Saurous}]{dillon_tensorflow_2017}
Dillon, J.~V., Langmore, I., Tran, D., {et~al.} 2017, {TensorFlow} {Distributions},  arXiv.
\newblock \url{http://arxiv.org/abs/1711.10604}

\bibitem[{Ellien {et~al.}(2021)Ellien, Slezak, Martinet, Durret, Adami, Gavazzi, Rabaça, Rocha, \& Pereira}]{ellien_dawis_2021}
Ellien, A., Slezak, E., Martinet, N., {et~al.} 2021, Astronomy \& Astrophysics, 649, A38, \dodoi{10.1051/0004-6361/202038419}

\bibitem[{Feldmeier {et~al.}(2004)Feldmeier, Mihos, Morrison, Harding, Kaib, \& Dubinski}]{feldmeier_deep_2004}
Feldmeier, J.~J., Mihos, J.~C., Morrison, H.~L., {et~al.} 2004, The Astrophysical Journal, 609, 617, \dodoi{10.1086/421313}

\bibitem[{Furnell {et~al.}(2021)Furnell, Collins, Kelvin, Baldry, James, Manolopoulou, Mann, Giles, Bermeo, Hilton, Wilkinson, Romer, Vergara, Bhargava, Stott, Mayers, \& Viana}]{furnell_growth_2021}
Furnell, K.~E., Collins, C.~A., Kelvin, L.~S., {et~al.} 2021, Monthly Notices of the Royal Astronomical Society, 502, 2419, \dodoi{10.1093/mnras/stab065}

\bibitem[{Geirhos {et~al.}(2020)Geirhos, Jacobsen, Michaelis, Zemel, Brendel, Bethge, \& Wichmann}]{geirhos_shortcut_2020}
Geirhos, R., Jacobsen, J.-H., Michaelis, C., {et~al.} 2020, Nature Machine Intelligence, 2, 665, \dodoi{10.1038/s42256-020-00257-z}

\bibitem[{Gladders \& Yee(2000)}]{gladders_new_2000}
Gladders, M.~D., \& Yee, H. K.~C. 2000, The Astronomical Journal, 120, 2148, \dodoi{10.1086/301557}

\bibitem[{Gonzalez {et~al.}(2005)Gonzalez, Zabludoff, \& Zaritsky}]{gonzalez_intracluster_2005}
Gonzalez, A.~H., Zabludoff, A.~I., \& Zaritsky, D. 2005, The Astrophysical Journal, 618, 195, \dodoi{10.1086/425896}

\bibitem[{Guennou {et~al.}(2012)Guennou, Adami, Rocha, Durret, Ulmer, Allam, Basa, Benoist, Biviano, Clowe, Gavazzi, Halliday, Ilbert, Johnston, Just, Kron, Kubo, Brun, Marshall, Mazure, Murphy, Pereira, Rabaça, Rostagni, Rudnick, Russeil, Schrabback, Slezak, Tucker, \& Zaritsky}]{guennou_intracluster_2012}
Guennou, L., Adami, C., Rocha, C.~D., {et~al.} 2012, Astronomy \& Astrophysics, 537, A64, \dodoi{10.1051/0004-6361/201117482}

\bibitem[{Han {et~al.}(2022)Han, Zou, Li, \& Chen}]{han_identifying_2022}
Han, Y., Zou, Z., Li, N., \& Chen, Y. 2022, Research in Astronomy and Astrophysics, 22, 085006, \dodoi{10.1088/1674-4527/ac7386}

\bibitem[{Hayat {et~al.}(2021)Hayat, Stein, Harrington, Lukić, \& Mustafa}]{hayat_self-supervised_2021}
Hayat, M.~A., Stein, G., Harrington, P., Lukić, Z., \& Mustafa, M. 2021, The Astrophysical Journal Letters, 911, L33, \dodoi{10.3847/2041-8213/abf2c7}

\bibitem[{He {et~al.}(2016)He, Zhang, Ren, \& Sun}]{he_deep_2016}
He, K., Zhang, X., Ren, S., \& Sun, J. 2016, in 2016 {IEEE} {Conference} on {Computer} {Vision} and {Pattern} {Recognition} ({CVPR}), 770--778, \dodoi{10.1109/CVPR.2016.90}

\bibitem[{Huang {et~al.}(2018)Huang, Leauthaud, Greene, Bundy, Lin, Tanaka, Miyazaki, \& Komiyama}]{huang_individual_2018}
Huang, S., Leauthaud, A., Greene, J.~E., {et~al.} 2018, Monthly Notices of the Royal Astronomical Society, 475, 3348, \dodoi{10.1093/mnras/stx3200}

\bibitem[{Huang {et~al.}(2020)Huang, Leauthaud, Hearin, Behroozi, Bradshaw, Ardila, Speagle, Tenneti, Bundy, Greene, Sifón, \& Bahcall}]{huang_weak_2020}
Huang, S., Leauthaud, A., Hearin, A., {et~al.} 2020, Monthly Notices of the Royal Astronomical Society, 492, 3685, \dodoi{10.1093/mnras/stz3314}

\bibitem[{Iodice {et~al.}(2020)Iodice, Spavone, Cattapan, Bannikova, Forbes, Rampazzo, Ciroi, Corsini, D’Ago, Oosterloo, Schipani, \& Capaccioli}]{iodice_vegas_2020}
Iodice, E., Spavone, M., Cattapan, A., {et~al.} 2020, Astronomy \& Astrophysics, 635, A3, \dodoi{10.1051/0004-6361/201936435}

\bibitem[{Ivezić {et~al.}(2019)Ivezić, Kahn, Tyson, Abel, Acosta, Allsman, Alonso, AlSayyad, Anderson, Andrew, Angel, Angeli, Ansari, Antilogus, Araujo, Armstrong, Arndt, Astier, Aubourg, Auza, Axelrod, Bard, Barr, Barrau, Bartlett, Bauer, Bauman, Baumont, Bechtol, Bechtol, Becker, Becla, Beldica, Bellavia, Bianco, Biswas, Blanc, Blazek, Blandford, Bloom, Bogart, Bond, Booth, Borgland, Borne, Bosch, Boutigny, Brackett, Bradshaw, Brandt, Brown, Bullock, Burchat, Burke, Cagnoli, Calabrese, Callahan, Callen, Carlin, Carlson, Chandrasekharan, Charles-Emerson, Chesley, Cheu, Chiang, Chiang, Chirino, Chow, Ciardi, Claver, Cohen-Tanugi, Cockrum, Coles, Connolly, Cook, Cooray, Covey, Cribbs, Cui, Cutri, Daly, Daniel, Daruich, Daubard, Daues, Dawson, Delgado, Dellapenna, Peyster, Val-Borro, Digel, Doherty, Dubois, Dubois-Felsmann, Durech, Economou, Eifler, Eracleous, Emmons, Neto, Ferguson, Figueroa, Fisher-Levine, Focke, Foss, Frank, Freemon, Gangler, Gawiser, Geary, Gee, Geha, Gessner, Gibson, Gilmore, Glanzman,
  Glick, Goldina, Goldstein, Goodenow, Graham, Gressler, Gris, Guy, Guyonnet, Haller, Harris, Hascall, Haupt, Hernandez, Herrmann, Hileman, Hoblitt, Hodgson, Hogan, Howard, Huang, Huffer, Ingraham, Innes, Jacoby, Jain, Jammes, Jee, Jenness, Jernigan, Jevremović, Johns, Johnson, Johnson, Jones, Juramy-Gilles, Jurić, Kalirai, Kallivayalil, Kalmbach, Kantor, Karst, Kasliwal, Kelly, Kessler, Kinnison, Kirkby, Knox, Kotov, Krabbendam, Krughoff, Kubánek, Kuczewski, Kulkarni, Ku, Kurita, Lage, Lambert, Lange, Langton, Guillou, Levine, Liang, Lim, Lintott, Long, Lopez, Lotz, Lupton, Lust, MacArthur, Mahabal, Mandelbaum, Markiewicz, Marsh, Marshall, Marshall, May, McKercher, McQueen, Meyers, Migliore, Miller, Mills, Miraval, Moeyens, Moolekamp, Monet, Moniez, Monkewitz, Montgomery, Morrison, Mueller, Muller, Arancibia, Neill, Newbry, Nief, Nomerotski, Nordby, O’Connor, Oliver, Olivier, Olsen, O’Mullane, Ortiz, Osier, Owen, Pain, Palecek, Parejko, Parsons, Pease, Peterson, Peterson, Petravick, Petrick, Petry,
  Pierfederici, Pietrowicz, Pike, Pinto, Plante, Plate, Plutchak, Price, Prouza, Radeka, Rajagopal, Rasmussen, Regnault, Reil, Reiss, Reuter, Ridgway, Riot, Ritz, Robinson, Roby, Roodman, Rosing, Roucelle, Rumore, Russo, Saha, Sassolas, Schalk, Schellart, Schindler, Schmidt, Schneider, Schneider, Schoening, Schumacher, Schwamb, Sebag, Selvy, Sembroski, Seppala, Serio, Serrano, Shaw, Shipsey, Sick, Silvestri, Slater, Smith, Smith, Sobhani, Soldahl, Storrie-Lombardi, Stover, Strauss, Street, Stubbs, Sullivan, Sweeney, Swinbank, Szalay, Takacs, Tether, Thaler, Thayer, Thomas, Thornton, Thukral, Tice, Trilling, Turri, Berg, Berk, Vetter, Virieux, Vucina, Wahl, Walkowicz, Walsh, Walter, Wang, Wang, Warner, Wiecha, Willman, Winters, Wittman, Wolff, Wood-Vasey, Wu, Xin, Yoachim, \& Zhan}]{ivezic_lsst_2019}
Ivezić, {\u Z}., Kahn, S.~M., Tyson, J.~A., {et~al.} 2019, The Astrophysical Journal, 873, 111, \dodoi{10.3847/1538-4357/ab042c}

\bibitem[{Jiménez-Teja \& Dupke(2016)}]{jimenez-teja_disentangling_2016}
Jiménez-Teja, Y., \& Dupke, R. 2016, The Astrophysical Journal, 820, 49, \dodoi{10.3847/0004-637X/820/1/49}

\bibitem[{Jiménez-Teja {et~al.}(2023)Jiménez-Teja, Dupke, Lopes, \& Vílchez}]{jimenez-teja_dissecting_2023}
Jiménez-Teja, Y., Dupke, R.~A., Lopes, P. A.~A., \& Vílchez, J.~M. 2023, Astronomy \& Astrophysics, 676, A39, \dodoi{10.1051/0004-6361/202346580}

\bibitem[{Jiménez-Teja {et~al.}(2018)Jiménez-Teja, Dupke, Benítez, Koekemoer, Zitrin, Umetsu, Ziegler, Frye, Ford, Bouwens, Bradley, Broadhurst, Coe, Donahue, Graves, Grillo, Infante, Jouvel, Kelson, Lahav, Lazkoz, Lemze, Maoz, Medezinski, Melchior, Meneghetti, Mercurio, Merten, Molino, Moustakas, Nonino, Ogaz, Riess, Rosati, Sayers, Seitz, \& Zheng}]{jimenez-teja_unveiling_2018}
Jiménez-Teja, Y., Dupke, R., Benítez, N., {et~al.} 2018, The Astrophysical Journal, 857, 79, \dodoi{10.3847/1538-4357/aab70f}

\bibitem[{Kingma \& Ba(2014)}]{kingma_adam_2014}
Kingma, D.~P., \& Ba, J. 2014, Adam: {A} {Method} for {Stochastic} {Optimization}, \dodoi{10.48550/arXiv.1412.6980}

\bibitem[{Kluge {et~al.}(2021)Kluge, Bender, Riffeser, Goessl, Hopp, Schmidt, \& Ries}]{kluge_photometric_2021}
Kluge, M., Bender, R., Riffeser, A., {et~al.} 2021, The Astrophysical Journal Supplement Series, 252, 27, \dodoi{10.3847/1538-4365/abcda6}

\bibitem[{Kluge {et~al.}(2024)Kluge, Hatch, Montes, Golden-Marx, Gonzalez, Cuillandre, Bolzonella, Lançon, Laureijs, Saifollahi, Schirmer, Stone, Boselli, Cantiello, Sorce, Marleau, Duc, Sola, Urbano, Ahad, Bahé, Bamford, Bellhouse, Buitrago, Dimauro, Durret, Ellien, Jimenez-Teja, Slezak, Aghanim, Altieri, Andreon, Auricchio, Baldi, Balestra, Bardelli, Bender, Bonino, Branchini, Brescia, Brinchmann, Camera, Candini, Capobianco, Carbone, Carretero, Casas, Castellano, Cavuoti, Cimatti, Congedo, Conselice, Conversi, Copin, Courbin, Courtois, Cropper, Da~Silva, Degaudenzi, Dinis, Duncan, Dupac, Dusini, Farina, Farrens, Ferriol, Fosalba, Frailis, Franceschi, Fumana, Galeotta, Garilli, Gillard, Gillis, Giocoli, Gómez-Alvarez, Granett, Grazian, Grupp, Guzzo, Haugan, Hoar, Hoekstra, Holmes, Hook, Hormuth, Hornstrup, Hudelot, Jahnke, Keihänen, Kermiche, Kiessling, Kitching, Kohley, Kubik, Kümmel, Kunz, Kurki-Suonio, Lahav, Ligori, Lilje, Lindholm, Lloro, Maiorano, Mansutti, Marggraf, Markovic, Martinet, Marulli,
  Massey, Maurogordato, McCracken, Medinaceli, Mei, Melchior, Mellier, Meneghetti, Merlin, Meylan, Moresco, Moscardini, Munari, Nichol, Niemi, Nightingale, Padilla, Paltani, Pasian, Pedersen, Percival, Pettorino, Pires, Polenta, Poncet, Popa, Pozzetti, Racca, Raison, Rebolo, Renzi, Rhodes, Riccio, Rix, Romelli, Roncarelli, Rossetti, Saglia, Sapone, Sartoris, Sauvage, Scaramella, Schneider, Schrabback, Secroun, Seidel, Seiffert, Serrano, Sirignano, Sirri, Skottfelt, Stanco, Tallada-Crespí, Taylor, Teplitz, Tereno, Toledo-Moreo, Torradeflot, Tutusaus, Valentijn, Valenziano, Vassallo, Kleijn, Veropalumbo, Wang, Weller, Williams, Zamorani, Zucca, Biviano, Burigana, De~Lucia, George, Scottez, Simon, Mora, Martín-Fleitas, Ruppin, \& Scott}]{kluge_euclid_2024}
Kluge, M., Hatch, N.~A., Montes, M., {et~al.} 2024, Euclid: {Early} {Release} {Observations} -- {The} intracluster light and intracluster globular clusters of the {Perseus} cluster,  arXiv.
\newblock \url{http://arxiv.org/abs/2405.13503}

\bibitem[{Krick \& Bernstein(2007)}]{krick_diffuse_2007}
Krick, J.~E., \& Bernstein, R.~A. 2007, The Astronomical Journal, 134, 466, \dodoi{10.1086/518787}

\bibitem[{Li {et~al.}(2022)Li, Huang, Leauthaud, Moustakas, Danieli, Greene, Abraham, Ardila, Kado-Fong, Lokhorst, Lupton, \& Price}]{li_reaching_2022}
Li, J., Huang, S., Leauthaud, A., {et~al.} 2022, Monthly Notices of the Royal Astronomical Society, 515, 5335, \dodoi{10.1093/mnras/stac2121}

\bibitem[{Loh \& Strauss(2006)}]{loh_bright_2006}
Loh, Y.-S., \& Strauss, M.~A. 2006, Monthly Notices of the Royal Astronomical Society, 366, 373, \dodoi{10.1111/j.1365-2966.2005.09714.x}

\bibitem[{Marini {et~al.}(2022)Marini, Borgani, Saro, Murante, Granato, Ragone-Figueroa, \& Taffoni}]{marini_machine_2022}
Marini, I., Borgani, S., Saro, A., {et~al.} 2022, Monthly Notices of the Royal Astronomical Society, 514, 3082, \dodoi{10.1093/mnras/stac1558}

\bibitem[{Martínez-Lombilla {et~al.}(2023)Martínez-Lombilla, Brough, Montes, Baena-Gallé, Akhlaghi, Infante-Sainz, Driver, Holwerda, Pimbblet, \& Robotham}]{martinez-lombilla_galaxy_2023}
Martínez-Lombilla, C., Brough, S., Montes, M., {et~al.} 2023, Monthly Notices of the Royal Astronomical Society, 518, 1195, \dodoi{10.1093/mnras/stac3119}

\bibitem[{Mihos(2019)}]{mihos_deep_2019}
Mihos, J.~C. 2019, Deep {Imaging} of {Diffuse} {Light} {Around} {Galaxies} and {Clusters}: {Progress} and {Challenges}, \dodoi{10.48550/arXiv.1909.09456}

\bibitem[{Mihos {et~al.}(2005)Mihos, Harding, Feldmeier, \& Morrison}]{mihos_diffuse_2005}
Mihos, J.~C., Harding, P., Feldmeier, J., \& Morrison, H. 2005, The Astrophysical Journal, 631, L41, \dodoi{10.1086/497030}

\bibitem[{Miyazaki {et~al.}(2018)Miyazaki, Komiyama, Kawanomoto, Doi, Furusawa, Hamana, Hayashi, Ikeda, Kamata, Karoji, Koike, Kurakami, Miyama, Morokuma, Nakata, Namikawa, Nakaya, Nariai, Obuchi, Oishi, Okada, Okura, Tait, Takata, Tanaka, Tanaka, Terai, Tomono, Uraguchi, Usuda, Utsumi, Yamada, Yamanoi, Aihara, Fujimori, Mineo, Miyatake, Oguri, Uchida, Tanaka, Yasuda, Takada, Murayama, Nishizawa, Sugiyama, Chiba, Futamase, Wang, Chen, Ho, Liaw, Chiu, Ho, Lai, Lee, Jeng, Iwamura, Armstrong, Bickerton, Bosch, Gunn, Lupton, Loomis, Price, Smith, Strauss, Turner, Suzuki, Miyazaki, Muramatsu, Yamamoto, Endo, Ezaki, Ito, Kawaguchi, Sofuku, Taniike, Akutsu, Dojo, Kasumi, Matsuda, Imoto, Miwa, Suzuki, Takeshi, \& Yokota}]{miyazaki_hyper_2018}
Miyazaki, S., Komiyama, Y., Kawanomoto, S., {et~al.} 2018, Publications of the Astronomical Society of Japan, 70, S1, \dodoi{10.1093/pasj/psx063}

\bibitem[{Montenegro-Taborda {et~al.}(2023)Montenegro-Taborda, Rodriguez-Gomez, Pillepich, Avila-Reese, Sales, Rodríguez-Puebla, \& Hernquist}]{montenegro-taborda_growth_2023}
Montenegro-Taborda, D., Rodriguez-Gomez, V., Pillepich, A., {et~al.} 2023, Monthly Notices of the Royal Astronomical Society, 521, 800, \dodoi{10.1093/mnras/stad586}

\bibitem[{Montes(2019)}]{montes_intracluster_2019}
Montes, M. 2019, The intracluster light and its role in galaxy evolution in clusters,  arXiv.
\newblock \url{http://arxiv.org/abs/1912.01616}

\bibitem[{Montes(2022)}]{montes_faint_2022}
---. 2022, Nature Astronomy, 6, 308, \dodoi{10.1038/s41550-022-01616-z}

\bibitem[{Montes {et~al.}(2021)Montes, Brough, Owers, \& Santucci}]{montes_buildup_2021}
Montes, M., Brough, S., Owers, M.~S., \& Santucci, G. 2021, The Astrophysical Journal, 910, 45, \dodoi{10.3847/1538-4357/abddb6}

\bibitem[{{Montes} \& {Trujillo}(2014)}]{montes_intracluster_2014}
{Montes}, M., \& {Trujillo}, I. 2014, \apj, 794, 137, \dodoi{10.1088/0004-637X/794/2/137}

\bibitem[{Montes \& Trujillo(2018)}]{montes_intracluster_2018}
Montes, M., \& Trujillo, I. 2018, Monthly Notices of the Royal Astronomical Society, 474, 917, \dodoi{10.1093/mnras/stx2847}

\bibitem[{Montes \& Trujillo(2022)}]{montes_new_2022}
---. 2022, The Astrophysical Journal Letters, 940, L51, \dodoi{10.3847/2041-8213/ac98c5}

\bibitem[{Morishita {et~al.}(2017)Morishita, Abramson, Treu, Schmidt, Vulcani, \& Wang}]{morishita_characterizing_2017}
Morishita, T., Abramson, L.~E., Treu, T., {et~al.} 2017, The Astrophysical Journal, 846, 139, \dodoi{10.3847/1538-4357/aa8403}

\bibitem[{Murante {et~al.}(2007)Murante, Giovalli, Gerhard, Arnaboldi, Borgani, \& Dolag}]{murante_importance_2007}
Murante, G., Giovalli, M., Gerhard, O., {et~al.} 2007, Monthly Notices of the Royal Astronomical Society, 377, 2, \dodoi{10.1111/j.1365-2966.2007.11568.x}

\bibitem[{Nishizawa {et~al.}(2020)Nishizawa, Hsieh, Tanaka, \& Takata}]{nishizawa_photometric_2020}
Nishizawa, A.~J., Hsieh, B.-C., Tanaka, M., \& Takata, T. 2020, Photometric {Redshifts} for the {Hyper} {Suprime}-{Cam} {Subaru} {Strategic} {Program} {Data} {Release} 2,  arXiv.
\newblock \url{http://arxiv.org/abs/2003.01511}

\bibitem[{Oguri(2014)}]{oguri_cluster_2014}
Oguri, M. 2014, Monthly Notices of the Royal Astronomical Society, 444, 147, \dodoi{10.1093/mnras/stu1446}

\bibitem[{Oguri {et~al.}(2018)Oguri, Lin, Lin, Nishizawa, More, More, Hsieh, Medezinski, Miyatake, Jian, Lin, Takada, Okabe, Speagle, Coupon, Leauthaud, Lupton, Miyazaki, Price, Tanaka, Chiu, Komiyama, Okura, Tanaka, \& Usuda}]{oguri_optically-selected_2018}
Oguri, M., Lin, Y.-T., Lin, S.-C., {et~al.} 2018, Publications of the Astronomical Society of Japan, 70, S20, \dodoi{10.1093/pasj/psx042}

\bibitem[{Oquab {et~al.}(2014)Oquab, Bottou, Laptev, \& Sivic}]{oquab_learning_2014}
Oquab, M., Bottou, L., Laptev, I., \& Sivic, J. 2014, in 2014 {IEEE} {Conference} on {Computer} {Vision} and {Pattern} {Recognition}, 1717--1724, \dodoi{10.1109/CVPR.2014.222}

\bibitem[{Pan \& Yang(2010)}]{pan_survey_2010}
Pan, S.~J., \& Yang, Q. 2010, IEEE Transactions on Knowledge and Data Engineering, 22, 1345, \dodoi{10.1109/TKDE.2009.191}

\bibitem[{Pearson {et~al.}(2019)Pearson, Wang, Trayford, Petrillo, \& Van Der~Tak}]{pearson_identifying_2019}
Pearson, W.~J., Wang, L., Trayford, J.~W., Petrillo, C.~E., \& Van Der~Tak, F. F.~S. 2019, Astronomy \& Astrophysics, 626, A49, \dodoi{10.1051/0004-6361/201935355}

\bibitem[{Presotto {et~al.}(2014)Presotto, Girardi, Nonino, Mercurio, Grillo, Rosati, Biviano, Annunziatella, Balestra, Cui, Sartoris, Lemze, Ascaso, Moustakas, Ford, Fritz, Czoske, Ettori, Kuchner, Lombardi, Maier, Medezinski, Molino, Scodeggio, Strazzullo, Tozzi, Ziegler, Bartelmann, Benitez, Bradley, Brescia, Broadhurst, Coe, Donahue, Gobat, Graves, Kelson, Koekemoer, Melchior, Meneghetti, Merten, Moustakas, Munari, Postman, Regős, Seitz, Umetsu, Zheng, \& Zitrin}]{presotto_intracluster_2014}
Presotto, V., Girardi, M., Nonino, M., {et~al.} 2014, Astronomy \& Astrophysics, 565, A126, \dodoi{10.1051/0004-6361/201323251}

\bibitem[{Proctor {et~al.}(2024)Proctor, Lagos, Ludlow, \& Robotham}]{proctor_identifying_2024}
Proctor, K.~L., Lagos, C. d.~P., Ludlow, A.~D., \& Robotham, A. S.~G. 2024, Monthly Notices of the Royal Astronomical Society, 527, 2624, \dodoi{10.1093/mnras/stad3341}

\bibitem[{Purcell {et~al.}(2007)Purcell, Bullock, \& Zentner}]{purcell_shredded_2007}
Purcell, C.~W., Bullock, J.~S., \& Zentner, A.~R. 2007, The Astrophysical Journal, 666, 20, \dodoi{10.1086/519787}

\bibitem[{Ragusa {et~al.}(2021)Ragusa, Spavone, Iodice, Brough, Raj, Paolillo, Cantiello, Forbes, Marca, D’Ago, Rampazzo, \& Schipani}]{ragusa_vegas_2021}
Ragusa, R., Spavone, M., Iodice, E., {et~al.} 2021, Astronomy \& Astrophysics, 651, A39, \dodoi{10.1051/0004-6361/202039921}

\bibitem[{Ragusa {et~al.}(2023)Ragusa, Iodice, Spavone, Montes, Forbes, Brough, Mirabile, Cantiello, Paolillo, \& Schipani}]{ragusa_does_2023}
Ragusa, R., Iodice, E., Spavone, M., {et~al.} 2023, Astronomy \& Astrophysics, 670, L20, \dodoi{10.1051/0004-6361/202245530}

\bibitem[{Robotham {et~al.}(2011)Robotham, Norberg, Driver, Baldry, Bamford, Hopkins, Liske, Loveday, Merson, Peacock, Brough, Cameron, Conselice, Croom, Frenk, Gunawardhana, Hill, Jones, Kelvin, Kuijken, Nichol, Parkinson, Pimbblet, Phillipps, Popescu, Prescott, Sharp, Sutherland, Taylor, Thomas, Tuffs, Van~Kampen, \& Wijesinghe}]{robotham_galaxy_2011}
Robotham, A. S.~G., Norberg, P., Driver, S.~P., {et~al.} 2011, Monthly Notices of the Royal Astronomical Society, 416, 2640, \dodoi{10.1111/j.1365-2966.2011.19217.x}

\bibitem[{Román {et~al.}(2020)Román, Trujillo, \& Montes}]{roman_galactic_2020}
Román, J., Trujillo, I., \& Montes, M. 2020, Astronomy \& Astrophysics, 644, A42, \dodoi{10.1051/0004-6361/201936111}

\bibitem[{Rudick {et~al.}(2006)Rudick, Mihos, \& McBride}]{rudick_formation_2006}
Rudick, C.~S., Mihos, J.~C., \& McBride, C. 2006, The Astrophysical Journal, 648, 936, \dodoi{10.1086/506176}

\bibitem[{Rudick {et~al.}(2011)Rudick, Mihos, \& McBride}]{rudick_quantity_2011}
Rudick, C.~S., Mihos, J.~C., \& McBride, C.~K. 2011, The Astrophysical Journal, 732, 48, \dodoi{10.1088/0004-637X/732/1/48}

\bibitem[{Selvaraju {et~al.}(2020)Selvaraju, Cogswell, Das, Vedantam, Parikh, \& Batra}]{selvaraju_grad-cam_2020}
Selvaraju, R.~R., Cogswell, M., Das, A., {et~al.} 2020, International Journal of Computer Vision, 128, 336, \dodoi{10.1007/s11263-019-01228-7}

\bibitem[{Semih~Kayhan \& van Gemert(2020)}]{semih_kayhan_translation_2020}
Semih~Kayhan, O., \& van Gemert, J.~C. 2020, in 2020 {IEEE}/{CVF} {Conference} on {Computer} {Vision} and {Pattern} {Recognition} ({CVPR}), 14262--14273, \dodoi{10.1109/CVPR42600.2020.01428}

\bibitem[{Spavone {et~al.}(2017)Spavone, Capaccioli, Napolitano, Iodice, Grado, Limatola, Cooper, Cantiello, Forbes, Paolillo, \& Schipani}]{spavone_vegas_2017}
Spavone, M., Capaccioli, M., Napolitano, N.~R., {et~al.} 2017, Astronomy \& Astrophysics, 603, A38, \dodoi{10.1051/0004-6361/201629111}

\bibitem[{Stein {et~al.}(2022)Stein, Blaum, Harrington, Medan, \& Lukić}]{stein_mining_2022}
Stein, G., Blaum, J., Harrington, P., Medan, T., \& Lukić, Z. 2022, The Astrophysical Journal, 932, 107, \dodoi{10.3847/1538-4357/ac6d63}

\bibitem[{Tanaka {et~al.}(2018)Tanaka, Coupon, Hsieh, Mineo, Nishizawa, Speagle, Furusawa, Miyazaki, \& Murayama}]{tanaka_photometric_2018}
Tanaka, M., Coupon, J., Hsieh, B.-C., {et~al.} 2018, Publications of the Astronomical Society of Japan, 70, S9, \dodoi{10.1093/pasj/psx077}

\bibitem[{Tang {et~al.}(2018)Tang, Lin, Cui, Kang, Wang, Contini, \& Yu}]{tang_investigation_2018}
Tang, L., Lin, W., Cui, W., {et~al.} 2018, The Astrophysical Journal, 859, 85, \dodoi{10.3847/1538-4357/aabd78}

\bibitem[{{Tensorflow Developers}(2023)}]{tensorflow_developers_tensorflow_2023}
{Tensorflow Developers}. 2023, {TensorFlow},  Zenodo, \dodoi{10.5281/zenodo.10126399}

\bibitem[{van~der Walt {et~al.}(2014)van~der Walt, Schönberger, Nunez-Iglesias, Boulogne, Warner, Yager, Gouillart, Yu, \& contributors}]{van_der_walt_scikit-image_2014}
van~der Walt, S., Schönberger, J.~L., Nunez-Iglesias, J., {et~al.} 2014, PeerJ, 2, e453, \dodoi{10.7717/peerj.453}

\bibitem[{Vazdekis {et~al.}(2016)Vazdekis, Koleva, Ricciardelli, Röck, \& Falcón-Barroso}]{vazdekis_uv-extended_2016}
Vazdekis, A., Koleva, M., Ricciardelli, E., Röck, B., \& Falcón-Barroso, J. 2016, Monthly Notices of the Royal Astronomical Society, 463, 3409, \dodoi{10.1093/mnras/stw2231}

\bibitem[{Vojtekova {et~al.}(2021)Vojtekova, Lieu, Valtchanov, Altieri, Old, Chen, \& Hroch}]{vojtekova_learning_2021}
Vojtekova, A., Lieu, M., Valtchanov, I., {et~al.} 2021, Monthly Notices of the Royal Astronomical Society, 503, 3204, \dodoi{10.1093/mnras/staa3567}

\bibitem[{Walmsley {et~al.}(2022)Walmsley, Scaife, Lintott, Lochner, Etsebeth, Géron, Dickinson, Fortson, Kruk, Masters, Mantha, \& Simmons}]{walmsley_practical_2022}
Walmsley, M., Scaife, A. M.~M., Lintott, C., {et~al.} 2022, Monthly Notices of the Royal Astronomical Society, 513, 1581, \dodoi{10.1093/mnras/stac525}

\bibitem[{Watkins {et~al.}(2024)Watkins, Kaviraj, Collins, Knapen, Kelvin, Duc, Román, \& Mihos}]{watkins_strategies_2024}
Watkins, A.~E., Kaviraj, S., Collins, C.~C., {et~al.} 2024, Monthly Notices of the Royal Astronomical Society, 528, 4289, \dodoi{10.1093/mnras/stae236}

\bibitem[{Zech {et~al.}(2018)Zech, Badgeley, Liu, Costa, Titano, \& Oermann}]{zech_variable_2018}
Zech, J.~R., Badgeley, M.~A., Liu, M., {et~al.} 2018, PLOS Medicine, 15, e1002683, \dodoi{10.1371/journal.pmed.1002683}

\bibitem[{Zhang {et~al.}(2011)Zhang, Andernach, Caretta, Reiprich, Böhringer, Puchwein, Sijacki, \& Girardi}]{zhang_hiflugcs_2011}
Zhang, Y.-Y., Andernach, H., Caretta, C.~A., {et~al.} 2011, Astronomy \& Astrophysics, 526, A105, \dodoi{10.1051/0004-6361/201015830}

\end{thebibliography}
\bibliographystyle{aasjournal}

%% This command is needed to show the entire author+affiliation list when
%% the collaboration and author truncation commands are used.  It has to
%% go at the end of the manuscript.
%\allauthors

%% Include this line if you are using the \added, \replaced, \deleted
%% commands to see a summary list of all changes at the end of the article.
%\listofchanges

\end{document}